\documentclass[apjl]{emulateapj}
\usepackage{comment}
\usepackage{ifthen}
\usepackage{amsmath}

\newcommand{\ang}{\text{\normalfont\AA}}

\newcommand{\ctbd}[1]{}

\newcommand{\lc}{light curve}
\newcommand{\lcs}{light curves}
\newcommand{\Lc}{Light curve}

\newcommand{\band}[1]{\ensuremath{#1}-band}

\newcommand{\kms}{\ensuremath{\rm km\,s^{-1}}}
\newcommand{\ms}{\ensuremath{\rm m\,s^{-1}}}

\newcommand{\gcmc}{\ensuremath{\rm g\,cm^{-3}}}
\newcommand{\ergscmsq}{\ensuremath{\rm erg\,s^{-1}\,cm^{-2}}}

\newcommand{\logg}{\ensuremath{\log{g}}}
\newcommand{\vsini}{\ensuremath{v \sin{i}}}
\newcommand{\feh}{\ensuremath{\rm [Fe/H]}}

\newcommand{\rsun}{\ensuremath{R_\sun}}
\newcommand{\msun}{\ensuremath{M_\sun}}
\newcommand{\lsun}{\ensuremath{L_\sun}}

\newcommand{\rstar}{\ensuremath{R_\star}}
\newcommand{\mstar}{\ensuremath{M_\star}}
\newcommand{\lstar}{\ensuremath{L_\star}}

\newcommand{\teffstar}{\ensuremath{T_{\rm eff\star}}}
\newcommand{\rhostar}{\ensuremath{\rho_\star}}
\newcommand{\loggstar}{\ensuremath{\log{g_{\star}}}}

\newcommand{\mearth}{\ensuremath{M_\earth}}

\newcommand{\rpl}{\ensuremath{R_{p}}}
\newcommand{\mpl}{\ensuremath{M_{p}}}

\newcommand{\rhopl}{\ensuremath{\rho_{p}}}

\newcommand{\arstar}{\ensuremath{a/\rstar}}
\newcommand{\zrstar}{\ensuremath{\zeta/\rstar}}

\newcommand{\rjup}{\ensuremath{R_{\rm J}}}
\newcommand{\mjup}{\ensuremath{M_{\rm J}}}

\newcommand{\rjuplong}{\ensuremath{R_{\rm Jup}}}
\newcommand{\mjuplong}{\ensuremath{M_{\rm Jup}}}

\newcommand{\rhojuplong}{\ensuremath{\rho_{\rm Jup}}}

\newcommand{\reffigl}[1]{Figure~\ref{fig:#1}}
\newcommand{\refsecl}[1]{\mbox{Section \ref{sec:#1}}}

\newcommand{\reftabl}[1]{Table~\ref{tab:#1}}

\newcommand{\hatcurCCra}{\ensuremath{06^{\mathrm h}16^{\mathrm m}26.90{\mathrm s}}}                                  
\newcommand{\hatcurCCdec}{\ensuremath{-22{\arcdeg}32{\arcmin}48.8{\arcsec}}}                                 
\newcommand{\hatcurCCtwomass}{2MASS~06162689-2232487}                  
\newcommand{\hatcurCCgsc}{GSC~6505-00217}                              
\newcommand{\hatcurCCapassmVshort}{\ensuremath{13.46}}                
\newcommand{\hatcurCCapassmV}{\ensuremath{13.459\pm0.017}}             
\newcommand{\hatcurCCapassmB}{\ensuremath{14.267\pm0.020}}             
\newcommand{\hatcurCCtwomassJmag}{\ensuremath{12.032\pm0.022}}         
\newcommand{\hatcurCCtwomassHmag}{\ensuremath{11.697\pm0.025}}         
\newcommand{\hatcurCCtwomassKmag}{\ensuremath{11.612\pm0.021}}         
\newcommand{\hatcurLCrprstar}{\ensuremath{0.1132\pm0.0029}}            
\newcommand{\hatcurLCimp}{\ensuremath{0.217_{-0.093}^{+0.074}}}        
\newcommand{\hatcurLCzeta}{\ensuremath{21.26\pm0.15}}                  
\newcommand{\hatcurLCdur}{\ensuremath{0.1053\pm0.0009}}                
\newcommand{\hatcurLCingdur}{\ensuremath{0.0112\pm0.0006}}             
\newcommand{\hatcurLCP}{\ensuremath{2.516729\pm0.000002}}              
\newcommand{\hatcurLCPprec}{\ensuremath{2.5167286}}                    
\newcommand{\hatcurLCPshort}{\ensuremath{2.5167}}                      
\newcommand{\hatcurLCT}{\ensuremath{2455808.34095\pm0.00036}}          
\newcommand{\hatcurSMEiteff}{\ensuremath{5403\pm50}}                   
\newcommand{\hatcurSMEizfeh}{\ensuremath{0.43\pm0.08}}                 
\newcommand{\hatcurSMEizfehshort}{\ensuremath{0.43}}                   
\newcommand{\hatcurSMEilogg}{\ensuremath{4.46\pm0.10}}                 
\newcommand{\hatcurSMEivsin}{\ensuremath{1.5\pm0.5}}                   
\newcommand{\hatcurSMEivmac}{\ensuremath{NULL}}                        
\newcommand{\hatcurSMEivmic}{\ensuremath{NULL}}                        
\newcommand{\hatcurSMEiiteff}{\ensuremath{5500\pm98}}                  
\newcommand{\hatcurSMEiizfeh}{\ensuremath{0.00\pm0.1}}                 
\newcommand{\hatcurSMEiizfehshort}{\ensuremath{0.00}}                  
\newcommand{\hatcurSMEiilogg}{\ensuremath{4.50\pm0.02}}                
\newcommand{\hatcurSMEiivsin}{\ensuremath{5.49\pm2.07}}                
\newcommand{\hatcurSMEiivmac}{\ensuremath{NULL}}                       
\newcommand{\hatcurSMEiivmic}{\ensuremath{NULL}}                       
\newcommand{\hatcurLBii}{\ensuremath{0.3518}}                          
\newcommand{\hatcurLBiii}{\ensuremath{0.2930}}                         
\newcommand{\hatcurLBir}{\ensuremath{0.4675}}                          
\newcommand{\hatcurLBiir}{\ensuremath{0.2624}}                         
\newcommand{\hatcurLBiR}{\ensuremath{0.4356}}                          
\newcommand{\hatcurLBiiR}{\ensuremath{0.2718}}                         
\newcommand{\hatcurISOmshort}{\ensuremath{1.00}}                       
\newcommand{\hatcurISOmlong}{\ensuremath{1.001\pm0.020}}               
\newcommand{\hatcurISOrshort}{\ensuremath{0.92}}                       
\newcommand{\hatcurISOrlong}{\ensuremath{0.925\pm0.022}}               
\newcommand{\hatcurISOlogg}{\ensuremath{4.51\pm0.02}}                  
\newcommand{\hatcurISOlum}{\ensuremath{0.65\pm0.05}}                   
\newcommand{\hatcurISOmv}{\ensuremath{5.38\pm0.09}}                    
\newcommand{\hatcurISOage}{\ensuremath{2.1\pm1.6}}                     
\newcommand{\hatcurISOMK}{\ensuremath{3.54\pm0.06}}                    
\newcommand{\hatcurRVK}{\ensuremath{197.5\pm3.2}}                      
\newcommand{\hatcurRVrk}{\ensuremath{0.029\pm0.039}}                   
\newcommand{\hatcurRVrh}{\ensuremath{-0.097\pm0.085}}                  
\newcommand{\hatcurRVk}{\ensuremath{0.003\pm0.005}}                    
\newcommand{\hatcurRVh}{\ensuremath{-0.010_{-0.022}^{+0.010}}}         
\newcommand{\hatcurRVjitterA}{\ensuremath{29.0}}                       
\newcommand{\hatcurRVjitterB}{\ensuremath{8.4}}                        
\newcommand{\hatcurRVjitterC}{\ensuremath{0.0}}                        
\newcommand{\hatcurRVjitterD}{\ensuremath{5.8}}                        
\newcommand{\hatcurRVeccen}{\ensuremath{0.013\pm0.016}}                
\newcommand{\hatcurRVomega}{\ensuremath{275\pm98}}                     
\newcommand{\hatcurPPi}{\ensuremath{88.5\pm0.6}}                       
\newcommand{\hatcurPPlogg}{\ensuremath{3.50\pm0.03}}                   
\newcommand{\hatcurPPar}{\ensuremath{8.43_{-0.22}^{+0.17}}}            
\newcommand{\hatcurPParel}{\ensuremath{0.0362\pm0.0002}}               
\newcommand{\hatcurPPrho}{\ensuremath{1.55\pm0.16}}                    
\newcommand{\hatcurPPmshort}{\ensuremath{1.32}}                        
\newcommand{\hatcurPPmlong}{\ensuremath{1.323\pm0.028}}                
\newcommand{\hatcurPPrshort}{\ensuremath{1.02}}                        
\newcommand{\hatcurPPrlong}{\ensuremath{1.020\pm0.037}}                
\newcommand{\hatcurPPmrcorr}{\ensuremath{0.07}}                        
\newcommand{\hatcurPPteff}{\ensuremath{1315\pm21}}                     
\newcommand{\hatcurPPtheta}{\ensuremath{0.094\pm0.004}}                
\newcommand{\hatcurPPfluxavg}{\ensuremath{6.74\pm0.44}}                
\newcommand{\hatcurPPfluxavgdim}{\ensuremath{8}}                       
\newcommand{\hatcurXdist}{\ensuremath{420\pm12}}                       
\newcommand{\hatcur}{HATS-4}
\newcommand{\hatcurb}{HATS-4b}

\newcommand{\hatcurSMEversion}{i}                                       
\newcommand{\hatcurSMEteff}{\ifthenelse{\equal{\hatcurSMEversion}{i}}{\hatcurSMEiteff}{\hatcurSMEiiteff}}
\newcommand{\hatcurSMEzfeh}{\ifthenelse{\equal{\hatcurSMEversion}{i}}{\hatcurSMEizfeh}{\hatcurSMEiizfeh}}
\newcommand{\hatcurSMEzfehshort}{\ifthenelse{\equal{\hatcurSMEversion}{i}}{\hatcurSMEizfehshort}{\hatcurSMEiizfehshort}}
\newcommand{\hatcurSMElogg}{\ifthenelse{\equal{\hatcurSMEversion}{i}}{\hatcurSMEilogg}{\hatcurSMEiilogg}}
\newcommand{\hatcurSMEvsin}{\ifthenelse{\equal{\hatcurSMEversion}{i}}{\hatcurSMEivsin}{\hatcurSMEiivsin}}
\newcommand{\hatcurSMEvmac}{\ifthenelse{\equal{\hatcurSMEversion}{i}}{\hatcurSMEivmac}{\hatcurSMEiivmac}}
\newcommand{\hatcurSMEvmic}{\ifthenelse{\equal{\hatcurSMEversion}{i}}{\hatcurSMEivmic}{\hatcurSMEiivmic}}

\newcommand{\hatcurSMEversioneccen}{i}                                       
\newcommand{\hatcurSMEteffeccen}{\ifthenelse{\equal{\hatcurSMEversioneccen}{i}}{\hatcurSMEiteffeccen}{\hatcurSMEiiteffeccen}}
\newcommand{\hatcurSMEzfeheccen}{\ifthenelse{\equal{\hatcurSMEversioneccen}{i}}{\hatcurSMEizfeheccen}{\hatcurSMEiizfeheccen}}
\newcommand{\hatcurSMEzfehshorteccen}{\ifthenelse{\equal{\hatcurSMEversioneccen}{i}}{\hatcurSMEizfehshorteccen}{\hatcurSMEiizfehshorteccen}}
\newcommand{\hatcurSMEloggeccen}{\ifthenelse{\equal{\hatcurSMEversioneccen}{i}}{\hatcurSMEiloggeccen}{\hatcurSMEiiloggeccen}}
\newcommand{\hatcurSMEvsineccen}{\ifthenelse{\equal{\hatcurSMEversioneccen}{i}}{\hatcurSMEivsineccen}{\hatcurSMEiivsineccen}}
\newcommand{\hatcurSMEvmaceccen}{\ifthenelse{\equal{\hatcurSMEversioneccen}{i}}{\hatcurSMEivmaceccen}{\hatcurSMEiivmaceccen}}
\newcommand{\hatcurSMEvmiceccen}{\ifthenelse{\equal{\hatcurSMEversioneccen}{i}}{\hatcurSMEivmiceccen}{\hatcurSMEiivmiceccen}}

\newcommand{\hatcurisoshort}{YY}
\newcommand{\hatcurisocite}{yi:2001}
\newcommand{\hatcurlumind}{\arstar}
\newcommand{\hatcurjhkfilset}{ESO}

\newboolean{emulateapj}
\setboolean{emulateapj}{true}

\newboolean{rvtablelong}
\setboolean{rvtablelong}{true}

\newboolean{astroph}
\setboolean{astroph}{true}

\shortauthors{Jord\'an et al.}
\shorttitle{\hatcur\lowercase{b}}
\ifthenelse{\boolean{emulateapj}}{
    \newcommand{\titledag}{$\dagger$}
}{
    \newcommand{\titledag}{\dagger}
}

\begin{document}

\title{\hatcur\lowercase{b}: A Dense Hot-Jupiter Transiting a Super Metal-Rich G
  star\altaffilmark{\titledag}}

\author{Andr\'es Jord\'an\altaffilmark{1},
  Rafael Brahm\altaffilmark{1}, 
G.~\'A.~Bakos\altaffilmark{2,$\star$,$\star\star$},
D.~Bayliss\altaffilmark{3},
K.~Penev\altaffilmark{2},
J.~D.~Hartman\altaffilmark{2},
G.~Zhou\altaffilmark{3}, 
L.~Mancini\altaffilmark{4},
M.~Mohler-Fischer\altaffilmark{4},
S.~Ciceri\altaffilmark{4},
B.~Sato\altaffilmark{5},
Z.~Csubry\altaffilmark{2},
M.~Rabus\altaffilmark{1}, 
V.~Suc\altaffilmark{1},
N.~Espinoza\altaffilmark{1}, 
W.~Bhatti\altaffilmark{2},
M.~de~Val~Borro\altaffilmark{2},
L.~Buchhave\altaffilmark{6,7},
B.~Cs\'ak\altaffilmark{4}, 
T.~Henning\altaffilmark{4},
B.~Schmidt\altaffilmark{3},
T.~G.~Tan\altaffilmark{8},
R.~W.~Noyes\altaffilmark{7},
B.~B\'eky\altaffilmark{7},
R.~P.~Butler\altaffilmark{9},
S.~Shectman\altaffilmark{10},
J.~Crane\altaffilmark{10},
I.~Thompson\altaffilmark{10},
A.~Williams\altaffilmark{11},
R.~Martin\altaffilmark{11},
C.~Contreras\altaffilmark{12},
J.~L\'az\'ar\altaffilmark{13},
I.~Papp\altaffilmark{13},
P.~S\'ari\altaffilmark{13}
}

\altaffiltext{1}{Instituto de Astrof\'isica, Facultad de F\'isica,
    Pontificia Universidad Cat\'olica de Chile, Av.\ Vicu\~na Mackenna
    4860, 7820436 Macul, Santiago, Chile}
\altaffiltext{2}{Department of Astrophysical Sciences,
    Princeton University, NJ 08544, USA}
\altaffiltext{3}{The Australian National University, Canberra,
    Australia}
\altaffiltext{4}{Max Planck Institute for Astronomy, Heidelberg,
    Germany}
\altaffiltext{5}{Department of Earth and Planetary Sciences, Tokyo 
Institute of Technology, 2-12-1 Ookayama, Meguro-ku, Tokyo 152- 8551}
\altaffiltext{6}{Niels Bohr Institute, Copenhagen University, Denmark}
\altaffiltext{7}{Harvard-Smithsonian Center for Astrophysics,
    Cambridge, MA, USA}
\altaffiltext{8}{Perth Exoplanet Survey Telescope, Perth, Australia}
\altaffiltext{9}{Department of Terrestrial Magnetism, Carnegie Institution
of Washington, 5241 Broad Branch Road NW, Washington, DC 20015-1305,
USA}
\altaffiltext{10}{The Observatories of the Carnegie Institution of
  Washington, 813 Santa Barbara Street, Pasadena, CA 91101, USA}
\altaffiltext{11}{Perth Observatory, Walnut Road, Bickley, Perth 6076, WA, Australia}
\altaffiltext{12}{Centre for Astrophysics \& Supercomputing, Swinburne University of
Technology, PO Box 218, Hawthorn, VIC 3122, Australia}
\altaffiltext{13}{Hungarian Astronomical Association, Budapest,
    Hungary}
\altaffiltext{$\star$}{Alfred P.~Sloan Research Fellow}
\altaffiltext{$\star\star$}{Packard Fellow}
\altaffiltext{\titledag}{
  The HATSouth network is operated by a collaboration consisting of
  Princeton University (PU), the Max Planck Institut f\"ur Astronomie
  (MPIA), and the Australian National University (ANU).  The station
  at Las Campanas Observatory (LCO) of the Carnegie Institution is
  operated by PU in conjunction with collaborators at the Pontificia
  Universidad Cat\'olica de Chile (PUC), the station at the High
  Energy Spectroscopic Survey (HESS) site is operated in conjunction
  with MPIA, and the station at Siding Spring Observatory (SSO) is
  operated jointly with ANU.
This paper includes data gathered with the 6.5 meter Magellan
Telescopes located at Las Campanas Observatory, Chile.
Based in part on data collected at Subaru Telescope, which is operated
by the National Astronomical Observatory of Japan, and on observations
made with the MPG/ESO 2.2\,m Telescope at the ESO Observatory in La
Silla. This paper uses observations obtained with facilities of the
Las Cumbres Observatory Global Telescope.
}


\begin{abstract}
  We report the discovery by the HATSouth survey of \hatcurb{}, an
  extrasolar planet transiting a V=\hatcurCCapassmVshort\ mag G star.
  \hatcurb{} has a period of $P\approx\hatcurLCPshort$\,d, mass of
  $\mpl \approx \hatcurPPmshort$\,\mjuplong, radius of $\rpl \approx
  \hatcurPPrshort$\,\rjuplong{} and density of $\rhopl =\hatcurPPrho$
  \gcmc $\approx 1.24 \rhojuplong$. The host star has a mass of
  \hatcurISOmshort\,\msun, a radius of \hatcurISOrshort\,\rsun{} and a
  very high metallicity \feh$=\hatcurSMEzfeh$.  \hatcurb{} is among
  the densest known planets with masses between 1--2 \mjup{} and is
  thus likely to have a significant content of heavy elements of the
  order of 75 \mearth{}. In this paper we present the data reduction,
  radial velocity measurement and stellar classification techniques
  adopted by the HATSouth survey for the CORALIE spectrograph. We also
  detail a technique to estimate simultaneously \vsini{} and
  macroturbulence using high resolution spectra.


\setcounter{footnote}{0}
\end{abstract}



\section{Introduction}
\label{sec:introduction}

Planets that, from our vantage point, eclipse their host star as they
orbit are termed transiting exoplanets (TEPs). Thanks to this
fortuitous orientation of their orbits they allow us to measure a
range of physical quantities that are not accessible for non-transiting
systems. TEPs allow characterization of their atmospheres through the
techniques of emission and transmission spectroscopy, the measurement
of the projected angle between the stellar and orbital spins through
the Rossiter-McLaughlin effect, the measurement of the ratio of
planetary to stellar size $\rpl/\rstar$ and, combined with radial
velocities (RVs), a measurement of the true mass without ambiguities
arising form the unknown inclination. Due to their being amenable to a
more comprehensive physical characterization, TEPs have had a large
impact on our rapidly evolving theories of planet formation and
evolution.

One of the most basic quantities one can measure for a transiting
planet is its bulk density. Coupled with models, the bulk density
allows insights into the planetary composition, illustrated in a
straightforward way by the fact that increasing the mass fraction of
heavy elements increases the bulk density. The sample of confirmed
transiting gas giants shows a large range in bulk densities. Part of
this diversity is due to a population of planets with inflated radii,
a phenomenon which is empirically found to appear above an
orbit-averaged stellar irradiation level of $\langle F \rangle \sim
2\times 10^{8}$ erg s$^{-1}$ cm$^{-2}$ \citep{demory:2011:radii_flux}
and whose cause is not yet understood. The diversity in bulk density
persists for planets with lower irradiation
\citep[e.g.,][]{miller:2011:cores}, implying a large range in inferred
core masses ($M_\mathrm{C}$) ranging from planets consistent with
having negligible cores \citep[e.g.,
HAT-P-18b;][]{hartman:2011:hat1819} up to much denser planets with
core masses substantially bigger than those for the giants in our
Solar System \citep[e.g., HD149026 with $M_\mathrm{C} \approx 67$\,
\mearth;][]{sato:2005:hd149}. The core masses, and more generally the
heavy element content of giants as compared to their host stars, are
very informative in the context of models of planet formation, and the
extreme values observed for some systems can prove to be a challenge.

A correlation between the metallicity of the stars and the heavy
element content of their planets was reported early on by assuming a
mechanism to produce the inflated radii dependent on the incident flux
\citep{guillot:2006:Z,burrows:2007:Z}.  More recently
\citet{miller:2011:cores} determined the heavy element masses for
giant planets for a sample with lower irradiation levels in order to
circumvent the uncertainties imposed by the unknown heating or
contraction-stalling mechanism that operates at high irradiation. They
find that giant planets around metal-rich stars have higher levels of
heavy elements, with a minimum level $\sim$ 10--15 \mearth{} in heavy
elements for stars with [Fe/H] $\lesssim$ 0.  The scatter around this
relation is high, and further understanding of the heavy element
content of exoplanets necessitates more systems, particularly at
extremes in stellar metallicity to probe the most extreme
environments. Moreover, recent studies based on a larger sample size
suggest that the correlation may not hold at all
\citep[e.g.,][]{zhou:2014:hats5}.
In this work we present a system with extreme stellar metallicity
discovered by the HATSouth survey: \hatcurb{}, a Jupiter radius planet
of mass $\hatcurPPmshort$\, \mjup{} orbiting a super metal-rich star.

The layout of the paper is as follows. In \refsecl{obs} we report the
detection of the photometric signal and the follow-up spectroscopic
and photometric observations of \hatcur{}.  In \refsecl{analysis} we
describe the analysis of the data, beginning with the determination of
the stellar parameters, continuing with a discussion of the methods
used to rule out non-planetary, false positive scenarios which could
mimic the photometric and spectroscopic observations, and finishing
with a description of our global modeling of the photometry and RVs.
Our findings are discussed in \refsecl{discussion}.


\section{Observations}
\label{sec:obs}

\subsection{Photometric detection}
\label{sec:detection}

\ifthenelse{\boolean{emulateapj}}{
    \begin{deluxetable*}{llrrr}
}{
    \begin{deluxetable}{llrrr}
}
\tablewidth{0pc}
\tabletypesize{\scriptsize}
\tablecaption{
    Summary of photometric observations
    \label{tab:photobs}
}
\tablehead{
    \multicolumn{1}{c}{Facility}          &
    \multicolumn{1}{c}{Date(s)}             &
    \multicolumn{1}{c}{Number of Images}\tablenotemark{a}      &
    \multicolumn{1}{c}{Cadence (s)}\tablenotemark{b}         &
    \multicolumn{1}{c}{Filter}            \\
    &
    &
    &
    &
}
\startdata
HS-1 & 2009 Dec--2011 Apr & 12007 & 299 & Sloan~$r$ \\
HS-2 & 2009 Sep--2010 Sep & 10709 & 294 & Sloan~$r$ \\
HS-3 & 2009 Dec--2011 Feb &  2268 & 300 & Sloan~$r$ \\
HS-4 & 2009 Sep--2010 Sep &  5331 & 293 & Sloan~$r$ \\
HS-5 & 2010 Jan--2011 May &  2708 & 293 & Sloan~$r$ \\
HS-6 & 2010 Apr--2010 Sep &    98 & 295 & Sloan~$r$ \\
FTS 2.0\,m/Spectral  & 2012 Oct 20 & 107 &  82 & Sloan~$i$ \\
Perth 0.6\,m/Andor  & 2012 Nov 09 & 122 & 131 & Sloan~$r$ \\
Swope 1.0\,m/Site3  & 2012 Dec 30 & 20 & 154 & Sloan~$i$ \\
PEST 0.3\,m  & 2013 Jan 21 & 125 & 130 & $R_{C}$ \\
[-1.5ex]
\enddata 
\tablenotetext{a}{
  Excludes images which were rejected as significant outliers in the
  fitting procedure.
}
\tablenotetext{b}{
  The mode time difference rounded to the nearest second between
  consecutive points in each light curve.  Due to visibility, weather,
  pauses for focusing, etc., none of the light curves have perfectly
  uniform time sampling.
}
\ifthenelse{\boolean{emulateapj}}{
    \end{deluxetable*}
}{
    \end{deluxetable}
}

\begin{figure}[!ht]
\plotone{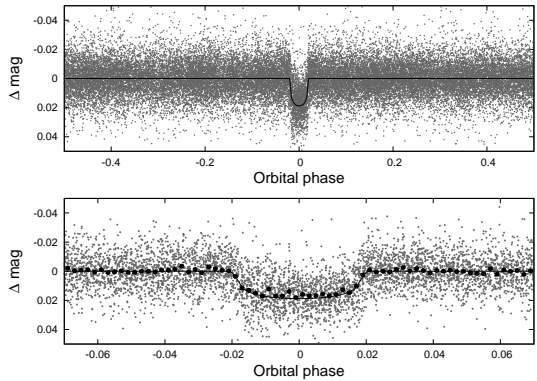}
\caption{
        Unbinned instrumental \band{r} \lc{} of \hatcur{} folded with
        the period $P = \hatcurLCPprec$\,days resulting from the global
        fit described in \refsecl{analysis}.  The solid line shows the
        best-fit transit model (see \refsecl{analysis}).  In the lower
        panel we zoom-in on the transit; the dark filled points here
        show the light curve binned in phase using a bin-size of 0.002.
\label{fig:hatsouth}}
\end{figure}

\hatcurb{} was first identified as a transiting exoplanet candidate
using a \lc{} constructed with 33,121 photometric measurements of its
host star \hatcur{} (also known as \hatcurCCtwomass{}; $\alpha =
\hatcurCCra$, $\delta = \hatcurCCdec$; J2000; V=\hatcurCCapassmV mag;
APASS DR5, \citealp{henden:2009}). The photometric measurements were
obtained with the full set of six HS units of the HATSouth project, a
global network of fully automated telescopes located at Las Campanas
Observatory (LCO) in Chile, at the location of the High Energy Spectroscopic Survey
(HESS) in Namibia, and at Siding Spring Observatory (SSO) in Australia.
A detailed description of HATSouth can be found in
\citet{bakos:2013:hats}.  Each of the HATSouth units consist of four
0.18 m f/2.8 Takahasi astrographs, each coupled with an Apogee U16M
Alta 4k$\times$4k CCD camera. Observations of 4 minutes are taken
through a Sloan $r$ filter, and covered the period between Sep 2009
and May 2011. The first six entries in Table~\ref{tab:photobs}
summarize the HATSouth discovery observations, and the differential
photometry data are presented in Table~\ref{tab:phfu}.
The large number of 33,121 photometric datapoints acquired for
\hatcur{} is remarkable, and this was possible due to the fact this
star was located at the intersection of two adjacent fields which were
monitored simultaneously by each of the two units at each site.
The majority of the datapoints were collected at the LCO stations HS-1
and HS-2, and very few observations were taken by the HS-6 unit in
Australia as its cameras were being serviced for a large fraction of
the time the relevant fields were being followed.

The photometric and candidate identification procedures adopted by
HATSouth are described in \citet{bakos:2013:hats} and
\citet{penev:2013:hats1}. Briefly, photometry is performed via
aperture photometry, and the resulting \lcs{} are detrended using
external parameter decorrelation (EPD) and trend filtering algorithm
(TFA, for a description of these procedures see the Appendix in
\citealp{bakos:2010:hat11} and references therein). The resulting
light curves are searched for transits using the Box-fitting Least
Squares (BLS) algorithm \citep{kovacs:2002:BLS}. The \hatcurb\
discovery light curve is shown in \reffigl{hatsouth}, folded with a
period of $P = \hatcurLCPprec$\,days. The transit evident in this
\lc{} triggered a follow-up process to confirm its planetary nature
that we describe below.

\subsection{Spectroscopy}
\label{sec:spec}

The process of spectroscopic confirmation for HATSouth candidates is
divided into steps of reconnaissance and high precision RV measurements
with high resolution spectrographs. \reftabl{specobssummary}
summarizes all the follow-up spectroscopic observations which we
obtained for \hatcur{}. The relative RVs and bisector span
measurements are presented in Table~\ref{tab:rvs}.
In the reconnaissance step, which utilizes spectrographs with a wide
range in resolutions, we are able to exclude most stellar binary false
positives. We obtained low and medium resolution reconnaissance
observations of \hatcur{} with the Wide Field Spectrograph WiFeS
\citep{dopita:2007:wifes} mounted on the ANU 2.3m telescope at SSO. A
flux calibrated spectrum at $R\equiv \lambda /  \Delta \lambda =3000$
provided an initial spectral classification for \hatcur{}, confirming
it was a dwarf and excluding false positive scenarios that involve
giants. Multiple-epoch observations at a higher resolution $R=7000$
allowed us to rule out variations in RV with amplitudes $> 1$ \kms{}
which would indicate the system is an eclipsing binary. Details of
WiFeS and the data reduction and analysis procedures adopted for
stellar parameter and RV measurement in the HATSouth survey have been
presented in \citet{bayliss:2013:hats3}.

After \hatcurb{} had passed the reconnaissance with WiFeS it was
scheduled to be observed on higher resolution facilities capable of
delivering high precision RVs. For \hatcurb{}, we obtained a total of
57 spectra with four such spectrographs as summarized in
Table~\ref{tab:specobssummary}. We now briefly describe the
observations and data reduction procedures adopted for each of these
spectrographs.

Twenty-four observations were obtained with the Planet Finder
Spectrograph \citep[PFS][]{crane:2010:PFS} mounted on the Magellan II
(Clay) telescope at LCO. We obtained two I$_2$-free templates on the
night of Dec 28 2012 UT with a slit width of 0\farcs3, while the
remaining observations were taken through an I$_2$ cell and a slit
width of 0\farcs5 on the nights of Dec 28-31 2012. A large number of
exposures with PFS were taken in-transit with the aim of measuring the
Rossiter-McLaughlin (R-M) effect (see \S~\ref{sec:macro}).
Exposures were read out using $2\times2$ binning on the CCD in order
to increase the signal-to-noise ratio (S/N), which was typically
$\approx$ 70 per resolution element for the I$_2$-free templates and
in the range 30--90 for the spectra taken through I$_2$. Reduction of the
PFS data was carried out using a custom pipeline and the RVs were
measured using the procedures described in \citet{butler:1996:I2}.

We also observed \hatcur{} with the High Dispersion Spectrograph
\citep[HDS,][]{noguchi:2002:hds} mounted on the Subaru telescope on
the nights Sep 19-22 2012, obtaining three I$_2$-free templates and
eight spectra taken through an I$_2$ cell. We used the KV370 filter
with a 0\farcs6$\times$2\arcsec slit, resulting in a resolution
$R=60,000$ and wavelength coverage 3500--6200 \ang.  Exposures were
read out using $2\times1$ binning on the CCD in order to increase the
S/N, which was typically $\approx$ 130 per resolution element for both
the I$_2$-free templates and the ones taken through I$_2$.  RVs were
measured using the procedure detailed in
\citet{sato:2002:I2hds,sato:2012:hatp38} which are in turn based on the
method of \citet{butler:1996:I2}. Bisector span measurements were
measured following \citet{bakos:2007:hd147}.

Finally, observations were also obtained with the FEROS spectrograph
\citep{kaufer:1998:feros} mounted on the ESO/MPG 2.2m telescope and
with the CORALIE spectrograph \citep{queloz:2001:coralie} mounted on the
Euler 1.2m telescope in La Silla. Descriptions of the data reduction
procedures for these spectrographs have been detailed in papers
reporting previous HATSouth discoveries
\citep{penev:2013:hats1,mohlerfischer:2013:hats2}. The CORALIE
procedures were only briefly described in \citet{penev:2013:hats1}. 
In the Appendix we describe in more detail reduction and RV measurement
procedures which we have developed and applied in the context of
HATSouth follow-up for CORALIE.

\hatcur{} is a metal-rich dwarf with low rotation, allowing in
principle very high precision RVs to be measured, but its relative
faint magnitude limits the precision that can be obtained in a single
measurement, especially for the spectrographs mounted on smaller
telescopes.

\ifthenelse{\boolean{emulateapj}}{
    \begin{deluxetable*}{llrrrrr}
}{
    \begin{deluxetable}{llrrr}
}
\tablewidth{0pc}
\tabletypesize{\scriptsize}
\tablecaption{
    Summary of spectroscopic observations\label{tab:specobssummary}
}
\tablehead{
    \multicolumn{1}{c}{Telescope/Instrument} &
    \multicolumn{1}{c}{Date Range}          &
    \multicolumn{1}{c}{Number of Observations} &
    \multicolumn{1}{c}{Resolution}          &
    \multicolumn{1}{c}{Observing Mode\tablenotemark{a}}          \\
}
\startdata
ANU 2.3\,m/WiFeS & 2012 May 8 & 1 & 3000 & RECON spectral class.\\
ANU 2.3\,m/WiFeS & 2012 May 10--12 & 3 & 7000 & RECON RVs\\
ESO/MPG 2.2\,m/FEROS & 2012 Aug 6 & 1 & 48000 & ThAr/HPRV \\
Euler 1.2\,m/CORALIE & 2012 Aug 21--25 & 3 & 60000 & ThAr/HPRV \\
Subaru 8.2\,m/HDS & 2012 Sep 19--22 & 8 & 60000 & I$_2$/HPRV \\
Subaru 8.2\,m/HDS & 2012 Sep 20 & 3 & 60000 & I$_2$-free template \\
Euler 1.2\,m/CORALIE & 2012 Nov 6--11 & 6 & 60000 & ThAr/HPRV \\
Magellan 6.5\,m/PFS & 2012 Dec 28 &3 & 127000 & I$_2$-free template \\
Magellan 6.5\,m/PFS & 2012 Dec 28--31 & 21 & 76000 & I$_2$/HPRV \\
ESO/MPG 2.2\,m/FEROS & 2012 Dec 29--31 & 6 & 48000 & ThAr/HPRV \\
ESO/MPG 2.2\,m/FEROS & 2013 Jan 22--27 & 4 & 48000 & ThAr/HPRV \\
ESO/MPG 2.2\,m/FEROS & 2013 Feb 24, 26 & 2 & 48000 & ThAr/HPRV \\
[-1.5ex]
\enddata 
\tablenotetext{a}{RECON spectral class. = reconnaissance spectral
  classification (see \S\ref{sec:spec}); RECON rvs = reconnaissance radial velocities
  (see \S\ref{sec:spec}); HPRV = high precision radial velocity,
  either via Iodine cell (I$_2$) or Thorium-Argon (ThAr) based measurements.
}
\ifthenelse{\boolean{emulateapj}}{
    \end{deluxetable*}
}{
    \end{deluxetable}
}

\begin{figure} [ht]
\plotone{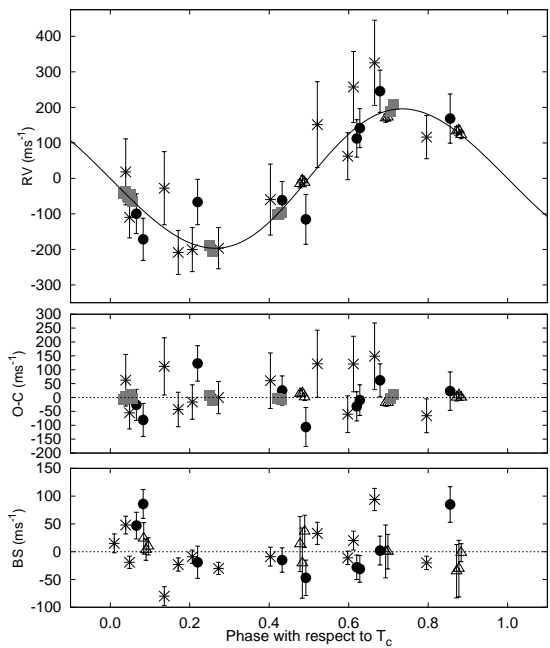}
\caption{
    {\em Top panel:} High-precision RV measurements for
    \hbox{\hatcur{}} from CORALIE (dark filled circles), Subaru/HDS
    (open triangles), FEROS (Xs) and PFS (gray-scale filled squares),
    together with our best-fit model.  Zero phase corresponds to the
    time of mid-transit.  The center-of-mass velocity has been
    subtracted.  {\em Second panel:} Velocity $O\!-\!C$ residuals from
    the best-fit model.  The error bars for each instrument have been
    inflated such that $\chi^2$ per degree of freedom is unity.  {\em
      Third panel:} Bisector spans (BS), with the mean value
    subtracted.  Note the different vertical scales of the panels.
\label{fig:rvbis}}
\end{figure}

\ifthenelse{\boolean{emulateapj}}{
    \begin{deluxetable*}{lrrrrrr}
}{
    \begin{deluxetable}{lrrrrrr}
}
\tablewidth{0pc}
\tablecaption{
    Relative radial velocities and bisector span measurements of
    \hatcur{}.
    \label{tab:rvs}
}
\tablehead{
    \colhead{BJD} & 
    \colhead{RV\tablenotemark{a}} & 
    \colhead{\ensuremath{\sigma_{\rm RV}}\tablenotemark{b}} & 
    \colhead{BS} & 
    \colhead{\ensuremath{\sigma_{\rm BS}}} & 
        \colhead{Phase} &
        \colhead{Instrument}\\
    \colhead{\hbox{(2\,454\,000$+$)}} & 
    \colhead{(\ms)} & 
    \colhead{(\ms)} &
    \colhead{(\ms)} &
    \colhead{} &
        \colhead{} &
        \colhead{}
}
\startdata
$ 2145.92503 $ & $   -25.66 $ & $   103.00 $ & $  -80.0 $ & $   17.0 $ & $   0.136 $ & FEROS \\
$ 2160.89156 $ & $  -172.31 $ & $    52.00 $ & $   86.0 $ & $   26.0 $ & $   0.083 $ & Coralie \\
$ 2161.92195 $ & $  -116.31 $ & $    64.00 $ & $  -47.0 $ & $   32.0 $ & $   0.492 $ & Coralie \\
$ 2164.90865 $ & $   244.69 $ & $    52.00 $ & $    2.0 $ & $   26.0 $ & $   0.679 $ & Coralie \\
$ 2190.11092 $ & $   172.54 $ & $     8.05 $ & $    0.5 $ & $   47.4 $ & $   0.693 $ & HDS \\
$ 2190.12564 $ & $   175.03 $ & $     7.41 $ & $    0.0 $ & $   31.1 $ & $   0.699 $ & HDS \\
$ 2191.09557\tablenotemark{c} $ & \nodata      & \nodata      & $   23.9 $ & $   28.3 $ & $   0.084 $ & HDS \\
$ 2191.11040\tablenotemark{c} $ & \nodata      & \nodata      & $    3.7 $ & $   19.1 $ & $   0.090 $ & HDS \\
$ 2191.12511\tablenotemark{c} $ & \nodata      & \nodata      & $    9.6 $ & $   15.4 $ & $   0.096 $ & HDS \\
$ 2192.08553 $ & $   -12.54 $ & $     7.67 $ & $   13.7 $ & $   49.6 $ & $   0.477 $ & HDS \\
$ 2192.10026 $ & $    -2.29 $ & $     8.80 $ & $  -20.6 $ & $   63.1 $ & $   0.483 $ & HDS \\
$ 2192.11498 $ & $    -8.90 $ & $     8.29 $ & $   36.6 $ & $   29.1 $ & $   0.489 $ & HDS \\
$ 2193.07700 $ & $   135.58 $ & $     9.52 $ & $  -35.1 $ & $   48.0 $ & $   0.871 $ & HDS \\
$ 2193.09173 $ & $   138.86 $ & $     7.95 $ & $  -30.5 $ & $   50.9 $ & $   0.877 $ & HDS \\
$ 2193.10646 $ & $   125.48 $ & $     6.77 $ & $   -1.8 $ & $   16.2 $ & $   0.883 $ & HDS \\
$ 2237.74591 $ & $   111.69 $ & $    44.00 $ & $  -28.0 $ & $   22.0 $ & $   0.620 $ & Coralie \\
$ 2238.86723 $ & $  -100.31 $ & $    48.00 $ & $   47.0 $ & $   24.0 $ & $   0.066 $ & Coralie \\
$ 2239.79009 $ & $   -62.31 $ & $    44.00 $ & $  -15.0 $ & $   22.0 $ & $   0.432 $ & Coralie \\
$ 2240.85420 $ & $   167.69 $ & $    63.00 $ & $   85.0 $ & $   32.0 $ & $   0.855 $ & Coralie \\
$ 2241.77140 $ & $   -67.31 $ & $    57.00 $ & $  -19.0 $ & $   29.0 $ & $   0.220 $ & Coralie \\
$ 2242.79877 $ & $   140.69 $ & $    47.00 $ & $  -31.0 $ & $   24.0 $ & $   0.628 $ & Coralie \\
$ 2289.66456 $ & $  -190.08 $ & $     2.94 $ & \nodata      & \nodata      & $   0.250 $ & PFS \\
$ 2289.68524 $ & $  -206.68 $ & $     3.77 $ & \nodata      & \nodata      & $   0.258 $ & PFS \\
$ 2290.57647 $ & $   259.34 $ & $   100.00 $ & $   20.0 $ & $   17.0 $ & $   0.612 $ & FEROS \\
$ 2290.71043 $ & $   327.34 $ & $   120.00 $ & $   94.0 $ & $   20.0 $ & $   0.665 $ & FEROS \\
$ 2290.81055 $ & $   187.02 $ & $     3.31 $ & \nodata      & \nodata      & $   0.705 $ & PFS \\
$ 2290.83195 $ & $   205.78 $ & $     3.74 $ & \nodata      & \nodata      & $   0.713 $ & PFS \\
$ 2291.53588 $\tablenotemark{d} & $    14.35 $ & $     5.55 $ & \nodata      & \nodata      & $   0.993 $ & PFS \\
$ 2291.54701 $\tablenotemark{d} & $    -5.57 $ & $     6.39 $ & \nodata      & \nodata      & $   0.998 $ & PFS \\
$ 2291.55792 $\tablenotemark{d} & $     8.57 $ & $     5.24 $ & \nodata      & \nodata      & $   0.002 $ & PFS \\
$ 2291.56891 $\tablenotemark{d} & $    -3.16 $ & $     5.16 $ & \nodata      & \nodata      & $   0.006 $ & PFS \\
$ 2291.57843 $\tablenotemark{d} & $    46.34 $ & $    96.00 $ & $   15.0 $ & $   17.0 $ & $   0.010 $ & FEROS \\
$ 2291.58011 $\tablenotemark{d} & $     5.48 $ & $     4.81 $ & \nodata      & \nodata      & $   0.011 $ & PFS \\
$ 2291.58849 $\tablenotemark{d} & $   -35.90 $ & $    10.38 $ & \nodata      & \nodata      & $   0.014 $ & PFS \\
$ 2291.60172 $\tablenotemark{d} & $   -17.41 $ & $     4.37 $ & \nodata      & \nodata      & $   0.019 $ & PFS \\
$ 2291.61378 $\tablenotemark{d} & $   -26.72 $ & $     4.99 $ & \nodata      & \nodata      & $   0.024 $ & PFS \\
$ 2291.62559 $\tablenotemark{d} & $   -27.07 $ & $     4.40 $ & \nodata      & \nodata      & $   0.029 $ & PFS \\
$ 2291.63700 $ & $   -44.46 $ & $     4.43 $ & \nodata      & \nodata      & $   0.033 $ & PFS \\
$ 2291.64813 $ & $   -39.03 $ & $     5.12 $ & \nodata      & \nodata      & $   0.038 $ & PFS \\
$ 2291.65109 $ & $    20.34 $ & $    93.00 $ & $   48.0 $ & $   16.0 $ & $   0.039 $ & FEROS \\
$ 2291.65909 $ & $   -52.96 $ & $     4.28 $ & \nodata      & \nodata      & $   0.042 $ & PFS \\
$ 2291.67013 $ & $   -54.64 $ & $     4.28 $ & \nodata      & \nodata      & $   0.047 $ & PFS \\
$ 2291.68108 $ & $   -46.23 $ & $     5.55 $ & \nodata      & \nodata      & $   0.051 $ & PFS \\
$ 2291.69219 $ & $   -66.57 $ & $     4.47 $ & \nodata      & \nodata      & $   0.055 $ & PFS \\
$ 2292.56735 $ & $   -57.66 $ & $   100.00 $ & $   -9.0 $ & $   17.0 $ & $   0.403 $ & FEROS \\
$ 2292.61509 $ & $  -103.30 $ & $     3.33 $ & \nodata      & \nodata      & $   0.422 $ & PFS \\
$ 2292.63646 $ & $   -96.78 $ & $     3.76 $ & \nodata      & \nodata      & $   0.430 $ & PFS \\
$ 2292.86458 $ & $   153.34 $ & $   121.00 $ & $   33.0 $ & $   20.0 $ & $   0.521 $ & FEROS \\
$ 2314.72311 $ & $  -198.66 $ & $    62.00 $ & $   -9.0 $ & $   12.0 $ & $   0.206 $ & FEROS \\
$ 2315.70746 $ & $    64.34 $ & $    66.00 $ & $  -11.0 $ & $   12.0 $ & $   0.598 $ & FEROS \\
$ 2318.72346 $ & $   118.34 $ & $    61.00 $ & $  -20.0 $ & $   12.0 $ & $   0.796 $ & FEROS \\
$ 2319.66753 $ & $  -206.66 $ & $    62.00 $ & $  -23.0 $ & $   12.0 $ & $   0.171 $ & FEROS \\
$ 2347.60681 $ & $  -194.66 $ & $    58.00 $ & $  -30.0 $ & $   11.0 $ & $   0.272 $ & FEROS \\
$ 2349.55979 $ & $  -107.66 $ & $    58.00 $ & $  -19.0 $ & $   11.0 $ & $   0.048 $ & FEROS \\

    [-1.5ex]
\enddata
\tablenotetext{a}{
        The zero-point of these velocities is arbitrary. An overall
        offset $\gamma_{\rm rel}$ fitted separately to the CORALIE
        and HDS velocities in \refsecl{analysis} has been
        subtracted.
}
\tablenotetext{b}{
        Internal errors excluding the component of
        astrophysical/instrumental jitter considered in
        \refsecl{analysis}.
}
\tablenotetext{c}{
        These HDS observations were taken without the iodine cell to
        be used as a template. RVs are not measured for these
        observations, but BS values are measured.
}
\tablenotetext{d}{
        These observations were obtained in transit and were excluded
        from our joint-fit analysis.
}
\ifthenelse{\boolean{emulateapj}}{
    \end{deluxetable*}
}{
    \end{deluxetable}
}

\subsection{Photometric follow-up observations}
\label{sec:phot}

\begin{figure}[!ht]
\plotone{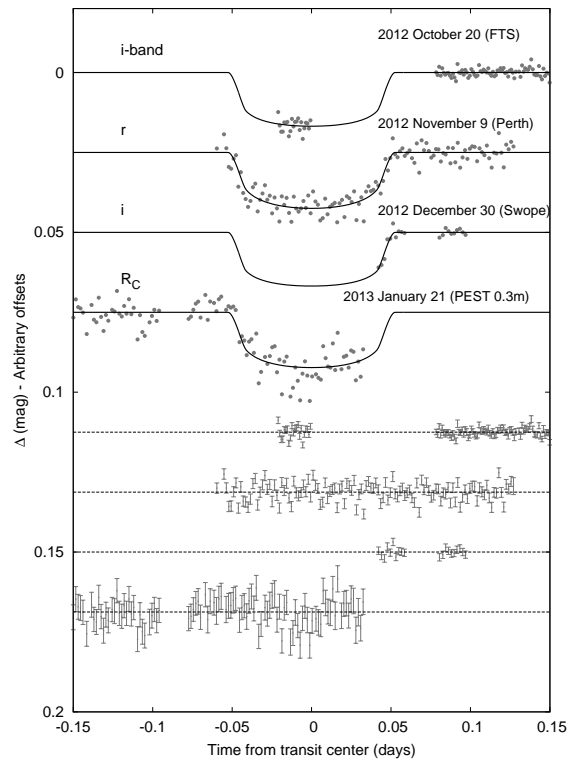}
\caption{
        Unbinned instrumental Sloan $r$- and \band{i}, and Cousins
        $R_{C}$ transit \lcs{} of \hatcur{}.  The dates and
        instruments used for each event are indicated.  The light
        curves have been detrended using the EPD process.  Curves
        after the first are shifted for clarity.  Our best fit is
        shown by the solid lines.  Residuals from the fits are
        displayed at the bottom, in the same order as the top curves.
\label{fig:lc}} \end{figure}

High precision photometric follow-up of \hatcur{}, necessary to
determine precise values of the orbital parameters and the planetary
radius, was performed using four facilities: the Spectral camera on
the 2m Faulkes Telescope South (FTS) at SSO, the SITe3 camera on the
Swope 1m telescope at LCO, the 0.3\,m Perth Exoplanet Survey Telescope
(PEST) and the Andor camera on the Perth 0.6m telescope. We observed
three partial and one full transit, the light curves are shown in
Figure~\ref{fig:lc} and the differential photometry for the light
curves is presented in Table~\ref{tab:phfu}. 

Details of the instruments, observation strategy, reduction and
photometric procedures for Spectral/FTS, SITe3/Swope and PEST can be
found in \citet{bayliss:2013:hats3}, \citet{penev:2013:hats1}, and
\citet{zhou:2014:lowmass}, respectively. This is the first time we
present data from the Perth 0.6\,m: on 2012 November 9 we monitored
the transit of \hatcur{} with the Andor iKon-L CCD camera coupled to
the 0.6\,m Perth-Lowell Automated Telescope at Perth Observatory,
Australia.  We used an \band{r} filter and exposure times of 100\,s.
Standard bias, dark, and flat-field corrections were performed using
the Perth-Lowell Automated Telescope automated reduction
pipeline. Aperture photometry was performed following the methods
described in \citet{bakos:2010:hat11} with modifications appropriate
for this facility.

\ifthenelse{\boolean{emulateapj}}{
        \begin{deluxetable*}{lrrrrr} }{
        \begin{deluxetable}{lrrrrr} 
    }
        \tablewidth{0pc}
        \tablecaption{Differential photometry of
        \hatcur\label{tab:phfu}} \tablehead{ \colhead{BJD} &
        \colhead{Mag\tablenotemark{a}} &
        \colhead{\ensuremath{\sigma_{\rm Mag}}} &
        \colhead{Mag(orig)\tablenotemark{b}} & \colhead{Filter} &
        \colhead{Instrument} \\ \colhead{\hbox{~~~~(2\,400\,000$+$)~~~~}}
        & \colhead{} & \colhead{} & \colhead{} & \colhead{} &
        \colhead{} } \startdata $ 55215.65143 $ & $  -0.00007 $ & $   0.00495 $ & $ \cdots $ & $ r$ &         HS\\
$ 55152.73411 $ & $  -0.00343 $ & $   0.00596 $ & $ \cdots $ & $ r$ &         HS\\
$ 55147.70097 $ & $   0.01909 $ & $   0.00654 $ & $ \cdots $ & $ r$ &         HS\\
$ 55167.83515 $ & $   0.01343 $ & $   0.00740 $ & $ \cdots $ & $ r$ &         HS\\
$ 55268.50492 $ & $  -0.00388 $ & $   0.00491 $ & $ \cdots $ & $ r$ &         HS\\
$ 55278.57207 $ & $  -0.00378 $ & $   0.00504 $ & $ \cdots $ & $ r$ &         HS\\
$ 55273.53947 $ & $   0.00252 $ & $   0.00470 $ & $ \cdots $ & $ r$ &         HS\\
$ 55215.65483 $ & $  -0.00724 $ & $   0.00493 $ & $ \cdots $ & $ r$ &         HS\\
$ 55152.73740 $ & $   0.00371 $ & $   0.00605 $ & $ \cdots $ & $ r$ &         HS\\
$ 55147.70431 $ & $  -0.00984 $ & $   0.00655 $ & $ \cdots $ & $ r$ &         HS\\

        [-1.5ex]
\enddata \tablenotetext{a}{
  The out-of-transit level has been subtracted. For the HATSouth light
  curve (rows with ``HS'' in the Instrument column), these magnitudes
  have been detrended using the EPD and TFA procedures prior to
  fitting a transit model to the light curve. Primarily as a result of
  this detrending, but also due to blending from neighbors, the
  apparent HATSouth transit depth is somewhat shallower than that of
  the true depth in the Sloan~$r$ filter (the apparent depth is
  92--100\% that of the true depth, depending on the field+detector
  combination). For the follow-up light curves (rows with an
  Instrument other than ``HS'') these magnitudes have been detrended
  with the EPD procedure, carried out simultaneously with the transit
  fit (the transit shape is preserved in this process).
}
\tablenotetext{b}{
        Raw magnitude values without application of the EPD
        procedure.  This is only reported for the follow-up light
        curves.
}
\tablecomments{
        This table is available in a machine-readable form in the
        online journal.  A portion is shown here for guidance regarding
        its form and content.
} \ifthenelse{\boolean{emulateapj}}{ \end{deluxetable*} }{ \end{deluxetable} }

\section{Analysis}
\label{sec:analysis}

\subsection{Stellar Parameter Estimation and Global Modeling of the
  Data}

The stellar parameters for \hatcurb{} were estimated using the Stellar
Parameter Classification (SPC) procedure \citep{buchhave:2012} using
the I$_2$-free templates obtained with PFS. The initial values of the
stellar parameters obtained with SPC were \teffstar$=5403\pm50$ K,
\logg$=4.46\pm0.1$, \feh$=0.43\pm 0.08$ and \vsini$=1.5\pm0.5$ \kms.

Joint global modeling of the available photometry and RV data was
performed following the procedure described in detail in
\citet{bakos:2010:hat11}. Light curves were modeled using the
expressions provided in \citet{mandel:2002} together with EPD and TFA
to account for systematic variations due to external parameters and
common trends of the ensemble of stars in a given set of
observations. The RV data was modeled with a Keplerian orbit following
the formalism of \citet{pal:2009:orbits}. The best fit parameters are
first estimated via a maximization of the likelihood using a downhill
simplex algorithm. This is followed by a Monte Carlo Markov Chain
(MCMC) analysis to estimate the posterior distribution of the
parameters.

Following \citet{sozzetti:2007} we use the values of \teffstar\
and \feh{} obtained from SPC together with the stellar mean density
\rhostar\ inferred from the transit to constrain mass, radius, age and
luminosity of \hatcur{} using the Yonsei-Yale (YY) stellar evolution
models \citep{yi:2001}. In each step of the MCMC run as part of the
global modeling, the current sample values for \teffstar, \rhostar\ and
\feh{} are used to interpolate from the YY isochrones values for mass,
radius, age and luminosity, building thus posterior distributions for
them and derived quantities such as \logg. This process usually
results in a more precise value of \logg{} than that inferred from
spectroscopy alone, and this new estimate for \logg{} can be fixed in
a new spectroscopic analysis to infer new values of \teffstar, \feh{} and
\vsini. The new values obtained are also used to update the limb
darkening parameters (taken from \citealp{claret:2004}) which are
needed to model the light curves and are dependent on stellar
parameters. This process is iterated until the isochrone-revised
\logg{} value is consistent within $1\sigma$ with the input
estimate. In the case of \hatcur{} the isochrone-revised value of
\logg{} was consistent with the original SPC value quoted above and
therefore no iterations were necessary.

\begin{figure}[!ht]
\plotone{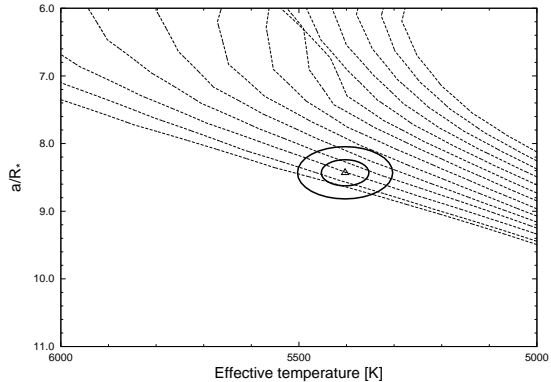}
\caption{
    Model isochrones from \cite{\hatcurisocite} for the measured
    metallicity of \hatcur, \feh = \hatcurSMEizfehshort, and ages of
    0.2\,Gyr, and 1 to 13\,Gyr in 1\,Gyr increments (left to right). 
    The adopted values of $\teffstar$ and \arstar\ are shown together
    with their 1$\sigma$ and 2$\sigma$ confidence ellipsoids.  
\label{fig:iso}}
\end{figure}

The stellar parameter estimates, along with 1$\sigma$ uncertainties
resulting from the process described above, are listed in
Table~\ref{tab:stellar}.  The inferred location of the star in a
diagram of \arstar{} versus \teffstar{} is shown in
Figure~\ref{fig:iso}. Finally, the global modeling estimates of the
geometric parameters relating to the light curves and the derived
physical parameters for \hatcurb{} are listed in
Table~\ref{tab:planetparam}.
We find that the planet has a mass of $\mpl \approx
\hatcurPPmshort$\,\mjup, and radius of $\rpl \approx
\hatcurPPrshort$\,\rjup, which results in a planet that is not
inflated and has a density of $\rhopl=\hatcurPPrho$ \gcmc.
Note that we allow the eccentricity to vary in the fit so as to allow
our uncertainty on this parameter to be propagated, but we find the
eccentricity to be consistent with that of a circular orbit.

\begin{figure}[!ht]
\plotone{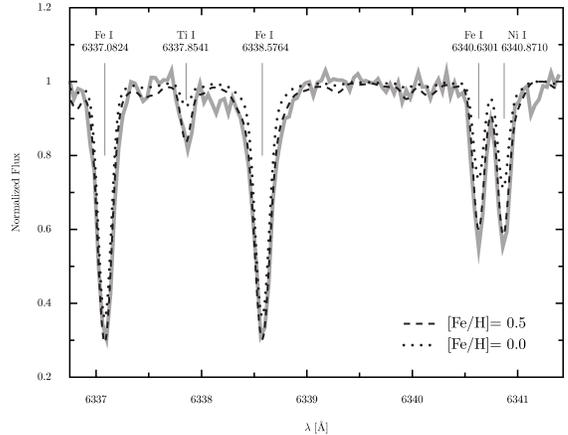}
\caption{
    Portion of an I$_2$-free spectrum of \hatcur{} taken with PFS
   and containing a few metallic lines identified in the figure along
   with their rest-frame vacuum wavelength values. The
    thick gray line connects the datapoints, while the dashed and dotted
    lines show PHOENIX models for values of
    \feh$=0.0,0.5$ as indicated, and with the rest of the
    stellar parameters fixed to the values listed in
    Table~\ref{tab:stellar}. It is clear that the very deep lines
    require a high \feh.
    \label{fig:feh}}
\end{figure}

One of the most remarkable properties of \hatcur{} is its high value
of \feh{}=\hatcurSMEzfeh. Of all the planet hosting stars listed in
\url{exoplanets.org} as of January 2014, only four stars have
\feh$>0.43$ dex, placing \hatcur{} among the 1\% most metal-rich
planet hosting stars.
To illustrate the high \feh{} of \hatcur{}, we show in
Figure~\ref{fig:feh} a portion of the PFS spectrum of \hatcur{} with a
few unsaturated metallic lines and overlay PHOENIX models
\citep{husser:2013:phoenix} for values of $\feh{}=0.0,0.5$ keeping the
other stellar parameters fixed to those listed in
Table~\ref{tab:stellar}. The figure clearly shows that high metal
enrichment values are needed to account for the very deep lines
present in the spectrum.

\subsection{Estimation of \vsini{} and Macroturbulence}
\label{sec:macro}

An initial analysis of the FEROS spectra suggested that \vsini{} was
$5\pm2$ \kms{}, and based on this information we scheduled
observations with PFS through a well-placed transit event to measure
the Rossiter-McLaughlin effect.
For a slowly rotating G star like \hatcur{} it is critical to
distinguish between rotational broadening (parametrized by $v \sin
i$) and broadening due to turbulence (parametrized by macroturbulence
$\zeta$) when modeling the spectra. Here we describe our procedure
for simultaneously measuring both of these parameters from the PFS data.

Both rotation and turbulence are large-scale motions on the surface of
the star, with the latter being generated by circulation and
convective cells that fill the surface. They result in a Doppler shift
of the radiation that emerges from the photosphere producing a
broadening of the spectral lines. For F and earlier type stars, the
rotational velocity is the principal source of line broadening.
On the other hand, for late-type stars like \hatcur{}, both velocity
phenomena are generally of similar magnitude \citep{gray:book}.
For these stars getting some hold on macroturbulence is necessary
because it broadens the lines in a comparable way to rotation, but the
R-M effect due to microturbulence is much smaller
\citep{albrecht:2012:obliq,boue:2013:arome}. Any line broadening
incorrectly attributed to rotation can thus adversely effect the
inferences on the relative alignment of stellar and orbital spins.

Disentangling the value of \vsini{} from $\zeta$ using the width of
the lines is a degenerate problem. At the velocity broadening expected
for late-type dwarfs (few km s$^{-1}$), the line profile is usually
dominated by the instrumental profile, and in addition it depends on
the atmospheric parameters of the star. The macroturbulence velocity
profile is very peaked while the rotational profile is much more
shallow, and if one is able to observe lines at very high resolution
and S/N, a Fourier analysis of single lines can separate
both effects \citep{gray:book}, a methodology that has been applied
for bright stars. For fainter systems like \hatcur{} this becomes
prohibitive, but one can exploit the many lines simultaneously
available in our spectra to constrain separately \vsini{} and $\zeta$.

The approach we followed is to select a large number of isolated,
unblended and unsaturated metal lines based on a PHOENIX synthetic
spectrum \citep{husser:2013:phoenix} with stellar parameters
corresponding to those measured with SPC. Then this set of lines for
the grid of synthetic models are convolved by the effects of \vsini\
and $\zeta$ for a grid of values which in our case was given by
$\{0.1,0.2,0.3,\ldots,7.8,7.9,8.0\}$ km s$^{-1}$ for both quantities.
The convolved models were generated following the radial-tangential
anisotropic macroturbulence model which assumes that a certain
fraction of the stellar material moves tangential to its surface and
the rest of the material moves along the stellar radius
\citep{gray:book}.  We assumed that both fractions are equal. To
compute the convolved line profile, a numerical disk integration was
performed at each wavelength including the macroturbulence Doppler
shift at each point on the disk and also the shift due to the stellar
rotation. We also include a linear limb darkening law in the disk
integration procedure using the coefficients of
\citet{claret:2011:ld}, performing a linear interpolation to obtain
the value at each wavelength.
Once we have the grid of convolved models, we compute the quantity
$X^2 = \sum_{k=1}^K\sum_{i} [m_i(\vsini,\zeta)-d_i]^2$, where
$m_i(\vsini,\zeta)$ is the convolved model and $d_i$ is the observed
spectrum. The index $k$ runs over the set of chosen lines while $i$
runs over the pixels belonging to each of the $K$ lines. The pair of
$(\hat{\vsini}, \hat{\zeta})$ values producing the lowest value of
$X^2$ is taken to be our estimates.
Based on the fact that $\zeta$ and \vsini{} only change the shape of
the spectral lines we normalize each line (both for the data and the
models) by its total absorbed flux, in order to reduce the effect of
systematic mismatches between data and models. Effectively, we are
thus fitting for the line shape and not its strength.  This
normalization also minimizes the effect of errors in the measured
\logg{} and \feh{} because in the case of non-saturated metal lines,
changes in these parameters mainly affect their strength but not their
shape or width.
To estimate the uncertainties on \vsini{} and $\zeta$, we applied a
bootstrap over the chosen metallic lines. That is, we selected a set
of lines with replacement from our list and re-run the fit obtaining a
sample of bootstrap estimates $\{\vsini_J,\zeta_J\}_{J=1}^B$ where $B$
is the number of bootstrap replications. From the bootstrap sample we
can estimate the parameter uncertainties by taking their dispersion.

We applied this method to the two I$_2$-free templates acquired with
PFS (see \S~\ref{sec:spec}). We chose 145 lines in the range between
5200--6500 \ang{} and due to the $2\times2$ binning used in the
observations we can sample each line with $\approx$ 10 pixels.
To get the instrumental response we used the calibration ThAr spectra
taken with the same slit we used in the observations.
First we computed a global spectral profile median combining all the
normalized emission lines of the ThAr spectrum. This global profile
was very well fitted by a Gaussian. We then fitted Gaussians to each
emission line to trace the instrumental response across our spectral
range. A clear trend was identified in each echelle order whereby the
spectral resolution increases from blue to red. This trend was
accounted for by fitting to each order the width of the ThAr lines as
a linear function of wavelength. We note that in observations of
stellar objects the slit is illuminated by the seeing profile, while
in a ThAr frame the slit is fully illuminated, leading in general to
different profiles for emission lines in the ThAr frames and stellar
objects in science observations. Given that the slit width we used is
small compared to the full-width at half-maximum of the seeing
profiles during our observations, using the ThAr frames to estimate
the instrumental response is an adequate approximation.

Our method yielded values of \vsini$=0.7\pm0.5$ km s$^{-1}$ and
$\zeta=1.3\pm0.4$ km s$^{-1}$ for \hatcur{}.
To test the sensitivity of our results on the assumed \teffstar, we
repeated the procedure but with a synthetic model with values of
\teffstar{} $\pm 1\sigma$ from the SPC values, obtaining results fully
consistent with the ones we quote.
As a validation procedure we measured \vsini{} and $\zeta$ for $\tau$
Ceti following the same procedure as for \hatcur. The stellar
parameters for $\tau$ Ceti were taken from
\citet{maldonado:2012:met}. We measured \vsini$=0.5\pm0.4$ km s$^{-1}$
and $\zeta=2.3\pm0.1$ km s$^{-1}$. The \vsini{} value is in agreement
with current studies which point towards $\tau$ Ceti having a rotation
axis close to perpendicular to the sky and the value for $\zeta$ is in
good agreement with the expected value for the spectral type of $\tau$
Ceti \citep{gray:book}.
We note that the total broadening by \vsini{} and $\zeta$ for
\hatcur{} is consistent with the value of \vsini{} estimated by SPC.

\begin{figure}[!ht]
\plotone{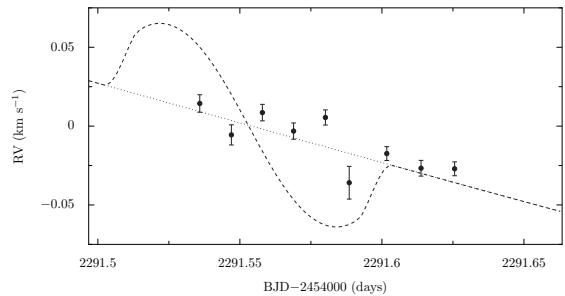}
\caption{
  Relative radial velocity (RV) measurements of \hatcur{} performed during transit
  using PFS. The dotted line shows the unperturbed orbit using the
  best fit orbital parameters determined in this work without
  considering the RVs shown. The dashed line illustrates what the
  Rossiter-McLaughlin effect would look like if the system was aligned
  and had $\vsini=5$ \kms.
\label{fig:RM}}
\end{figure}

In Figure~\ref{fig:RM} we show the RVs during transit measured using
PFS on the night of Dec 30 2012 (UT). Along with the data, we show the
expected behavior of the RVs without considering the
Rossiter-McLaughlin (R-M) effect, and the expected behavior assuming
the system was aligned and \vsini{} had the value of 5 \kms{}
initially estimated from the FEROS data. The R-M effect was calculated
for RVs measured with an I$_2$ cell using the ARoME library
\citep{boue:2013:arome}. It is clear that with \vsini$=5$ \kms{} the
precision afforded by PFS would have allowed us to measure the
relative projected angle between the stellar and orbital spins, but
with the actual value of \vsini$=0.5\pm0.4$ \kms{} the R-M effect is
barely noticeable and thus no constraints are possible.

\subsection{Excluding blend scenarios}

To rule out the possibility that \hatcur\ is a blended stellar
eclipsing binary system we conducted a blend analysis as described in
\cite{hartman:2011:hat3233}. This analysis attempts to model the
available light curves, photometry calibrated to an absolute scale,
and spectroscopically determined stellar atmospheric parameters, using
combinations of stars with parameters constrained to lie on the
\citet{girardi:2000} evolutionary tracks. We find that the data are
best described by a planet transiting a star. Moreover, the only
non-planetary blend models that cannot be rejected with greater than
$5\sigma$ confidence, based on the photometry alone, would have easily
been rejected as double-lined spectroscopic binary systems, or would
have exhibited very large ($\sim 1$\,\kms) RV and/or bisector-span
variations. Such large bisector-span variations can be ruled out based
on the Subaru/HDS measurements which show an RMS scatter of only $\sim
22$\,\ms.


\begin{deluxetable*}{lrl}
\tablewidth{0pc}
\tabletypesize{\scriptsize}
\tablecaption{
    Stellar parameters for \hatcur{}
    \label{tab:stellar}
}
\tablehead{
    \multicolumn{1}{c}{~~~~~~~~Parameter~~~~~~~~}   &
    \multicolumn{1}{c}{Value} &
    \multicolumn{1}{c}{Source}
}
\startdata
\noalign{\vskip -3pt}
\sidehead{Spectroscopic properties}
~~~~$\teffstar$ (K)\dotfill         &  \hatcurSMEteff & SPC\tablenotemark{a}\\
~~~~$\feh$\dotfill                  &  \hatcurSMEzfeh & SPC                 \\
~~~~$\vsini$ (\kms)\dotfill         &  $0.7\pm0.5$ & This work \\
~~~~$\zeta$ (\kms)\tablenotemark{b}\dotfill          &  $1.3\pm0.4$ & This work \\
\sidehead{Photometric properties}
~~~~$V$ (mag)\dotfill               &  \hatcurCCapassmV & APASS                \\
~~~~$B$ (mag)\dotfill               &  \hatcurCCapassmB & APASS                \\
~~~~$J$ (mag)\dotfill               &  \hatcurCCtwomassJmag & 2MASS           \\
~~~~$H$ (mag)\dotfill               &  \hatcurCCtwomassHmag & 2MASS           \\
~~~~$K_s$ (mag)\dotfill             &  \hatcurCCtwomassKmag & 2MASS           \\
\sidehead{Derived properties}
~~~~$\mstar$ ($\msun$)\dotfill      &  \hatcurISOmlong & \hatcurisoshort+\hatcurlumind+SPC \tablenotemark{c}\\
~~~~$\rstar$ ($\rsun$)\dotfill      &  \hatcurISOrlong & \hatcurisoshort+\hatcurlumind+SPC      \\
~~~~$\loggstar$ (cgs)\dotfill       &  \hatcurISOlogg & \hatcurisoshort+\hatcurlumind+SPC         \\
~~~~$\lstar$ ($\lsun$)\dotfill      &  \hatcurISOlum & \hatcurisoshort+\hatcurlumind+SPC         \\
~~~~$M_V$ (mag)\dotfill             &  \hatcurISOmv & \hatcurisoshort+\hatcurlumind+SPC         \\
~~~~$M_K$ (mag,\hatcurjhkfilset)\dotfill &  \hatcurISOMK & \hatcurisoshort+\hatcurlumind+SPC         \\
~~~~Age (Gyr)\dotfill               &  \hatcurISOage  & \hatcurisoshort+\hatcurlumind+SPC         \\
~~~~Distance (pc)\dotfill           &  \hatcurXdist   & \hatcurisoshort+\hatcurlumind+SPC\\
[-1.5ex]
\enddata
\tablenotetext{a}{
  SPC = ``Spectral Parameter Classification'' procedure for the
  measurement of stellar parameters from high-resolution spectra
  \citep{buchhave:2012}.  These parameters rely primarily on SPC, but
  have a small dependence also on the iterative analysis incorporating
  the isochrone search and global modeling of the data, as described
  in the text.
}
\tablenotetext{b}{
 $\zeta$ is the macroturbulence velocity estimated under the
 assumption that the tangential and radial components are equal (see \S\ref{sec:macro})
}
\tablenotetext{c}{
    \hatcurisoshort+\hatcurlumind+SPC = Based on the \hatcurisoshort\
    isochrones \citep{\hatcurisocite}, \hatcurlumind\ as a luminosity
    indicator, and the SPC results.
}
\end{deluxetable*}

\begin{deluxetable*}{lr}
\tabletypesize{\scriptsize}
\tablecaption{Orbital and planetary parameters\label{tab:planetparam}}
\tablehead{
        \multicolumn{1}{c}{~~~~~~~~~~~~~~~Parameter~~~~~~~~~~~~~~~} &
        \multicolumn{1}{c}{Value} \\
}
\startdata
\noalign{\vskip -3pt}
\sidehead{\Lc{} parameters}
~~~$P$ (days)             \dotfill    & $\hatcurLCP$ \\
~~~$T_c$ (${\rm BJD}$)    
      \tablenotemark{a}   \dotfill    & $\hatcurLCT$ \\
~~~$T_{14}$ (days)
      \tablenotemark{a}   \dotfill    & $\hatcurLCdur$ \\
~~~$T_{12} = T_{34}$ (days)
      \tablenotemark{a}   \dotfill    & $\hatcurLCingdur$ \\
~~~$\arstar$              \dotfill    & $\hatcurPPar$ \\
~~~$\zrstar$\tablenotemark{b}              \dotfill    & $\hatcurLCzeta$ \\
~~~$\rpl/\rstar$          \dotfill    & $\hatcurLCrprstar$ \\
~~~$b \equiv a \cos i/\rstar$
                          \dotfill    & $\hatcurLCimp$ \\
~~~$i$ (deg)              \dotfill    & $\hatcurPPi$ \\

\sidehead{Limb-darkening coefficients \tablenotemark{c}}
~~~$a_r$ (linear term)   \dotfill    & $\hatcurLBir$ \\
~~~$b_r$ (quadratic term) \dotfill    & $\hatcurLBiir$ \\
~~~$a_R$                 \dotfill    & $\hatcurLBiR$ \\
~~~$b_R$                 \dotfill    & $\hatcurLBiiR$ \\
~~~$a_i$                 \dotfill    & $\hatcurLBii$ \\
~~~$b_i$                  \dotfill    & $\hatcurLBiii$ \\

\sidehead{RV parameters}
~~~$K$ (\ms)              \dotfill    & $\hatcurRVK$ \\
~~~$\sqrt{e}\cos\omega$ 
                          \dotfill    & $\hatcurRVrk$ \\
~~~$\sqrt{e}\sin\omega$
                          \dotfill    & $\hatcurRVrh$ \\
~~~$e\cos\omega$ 
                          \dotfill    & $\hatcurRVk$ \\
~~~$e\sin\omega$
                          \dotfill    & $\hatcurRVh$ \\
~~~$e$                    \dotfill    & $\hatcurRVeccen$ \\
~~~$\omega$                    \dotfill    & $\hatcurRVomega$ \\
~~~CORALIE RV jitter (\ms)\tablenotemark{d}        
                          \dotfill    & $\hatcurRVjitterA$ \\
~~~HDS RV jitter (\ms)        
                          \dotfill    & $\hatcurRVjitterB$ \\
~~~FEROS RV jitter (\ms)        
                          \dotfill    & $\hatcurRVjitterC$ \\
~~~PFS RV jitter (\ms)        
                          \dotfill    & $\hatcurRVjitterD$ \\

\sidehead{Planetary parameters}
~~~$\mpl$ ($\mjup$)       \dotfill    & $\hatcurPPmlong$ \\
~~~$\rpl$ ($\rjup$)       \dotfill    & $\hatcurPPrlong$ \\
~~~$C(\mpl,\rpl)$
    \tablenotemark{e}     \dotfill    & $\hatcurPPmrcorr$ \\
~~~$\rhopl$ (\gcmc)       \dotfill    & $\hatcurPPrho$ \\
~~~$\log g_p$ (cgs)       \dotfill    & $\hatcurPPlogg$ \\
~~~$a$ (AU)               \dotfill    & $\hatcurPParel$ \\
~~~$T_{\rm eq}$ (K)       \dotfill    & $\hatcurPPteff$ \\
~~~$\Theta$\tablenotemark{f}\dotfill  & $\hatcurPPtheta$ \\
~~~$\langle F \rangle$ ($10^{\hatcurPPfluxavgdim}$\ergscmsq) 
\tablenotemark{g}         \dotfill    & $\hatcurPPfluxavg$ \\
[-1.5ex]
\enddata
\tablenotetext{a}{
    \ensuremath{T_c}: Reference epoch of mid transit that minimizes the
    correlation with the orbital period. BJD is calculated from UTC.
    \ensuremath{T_{14}}: total transit duration, time between first to
    last contact;
    \ensuremath{T_{12}=T_{34}}: ingress/egress time, time between first
    and second, or third and fourth contact.
}
\tablenotetext{b}{
    Reciprocal of the half duration of the transit used as a jump
    parameter in our MCMC analysis in place of $\arstar$. It is
    related to $\arstar$ by the expression $\zrstar = \arstar
    (2\pi(1+e\sin \omega))/(P \sqrt{1 - b^{2}}\sqrt{1-e^{2}})$
    \citep{bakos:2010:hat11}.
}
\tablenotetext{c}{
        Values for a quadratic law given separately for the Sloan~$g$,
        $r$, and $i$ filters.  These values were adopted from the
        tabulations by \cite{claret:2004} according to the
        spectroscopic (SPC) parameters listed in \reftabl{stellar}.
}
\tablenotetext{d}{
    This jitter was added in quadrature to the RV uncertainties for
    each instrument such that $\chi^{2}/{\rm dof} = 1$ for the
    observations from that instrument.
}
\tablenotetext{e}{
    Correlation coefficient between the planetary mass \mpl\ and radius
    \rpl.
}
\tablenotetext{f}{
    The Safronov number is given by $\Theta = \frac{1}{2}(V_{\rm
    esc}/V_{\rm orb})^2 = (a/\rpl)(\mpl / \mstar )$
    \citep[see][]{hansen:2007}.
}
\tablenotetext{g}{
    Incoming flux per unit surface area, averaged over the orbit.
}
\end{deluxetable*}

\section{Discussion}
\label{sec:discussion}

\begin{figure}[!ht]
\plotone{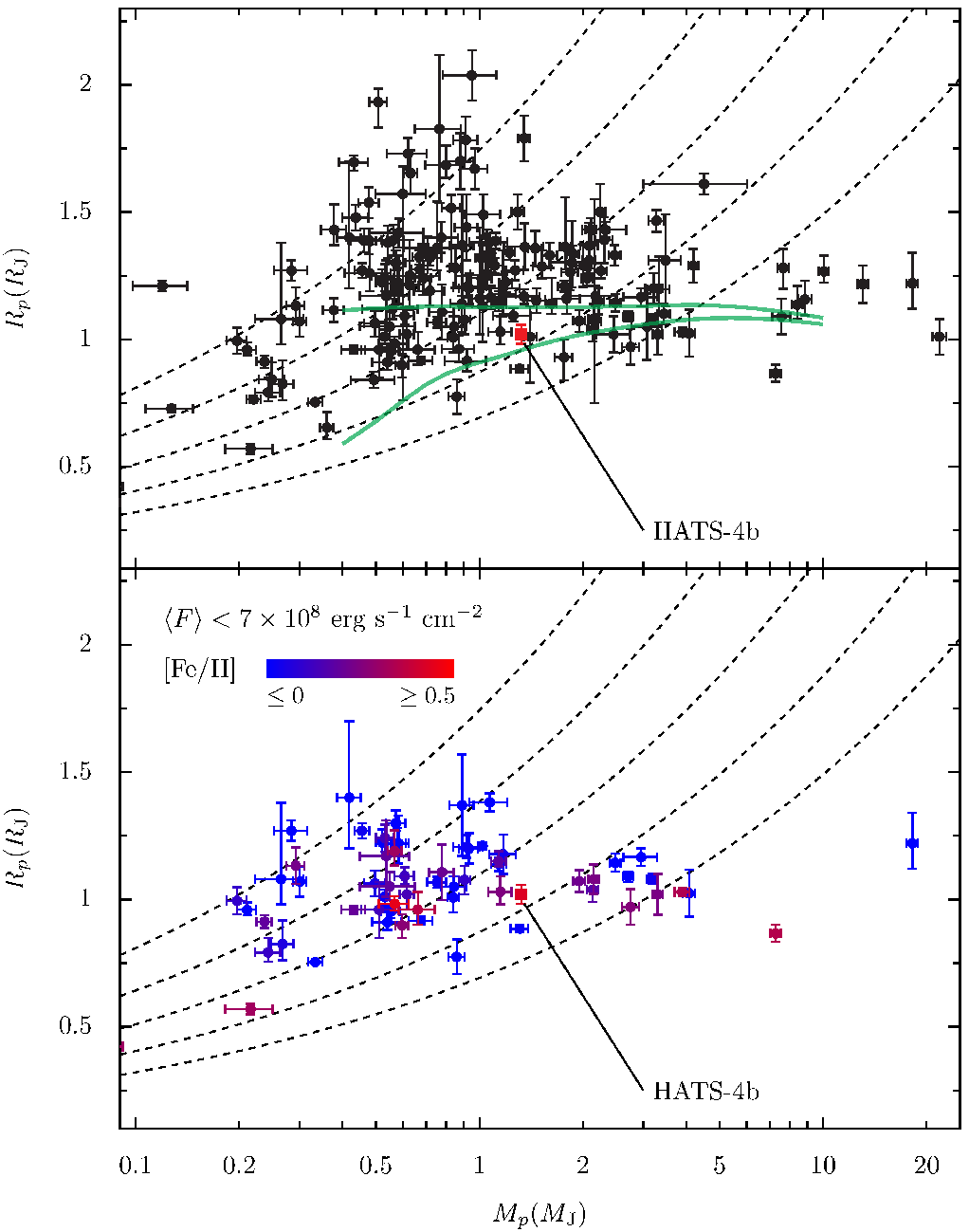}
\caption{
{\em Top panel:} 
Mass-radius diagram for a large sample of known TEPs (black circles). \hatcurb{} is indicated by the
red square. Overlaid as thick green lines are
\citet{fortney:2007:core_models} isochrones interpolated to the solar
equivalent semi-major axis of \hatcurb{} for an age of 4.5 Gyr and
core masses of 0 (upper) and 100 (lower) \mearth. We also overlay
isodensity lines for $\rho_p = \{0.25, 0.5, 1, 2, 4\}$ gr cm$^{-3}$
(dashed curves). Data taken from {\tt exoplanets.org} on Jan 9 2014,
adding HATS-3b \citep{bayliss:2013:hats3} and HATS-5b \citep{zhou:2014:hats5}.
{\em Bottom panel}: Same as top panel but restricted to planets with
incoming flux per unit surface area  (averaged over the orbit) $\langle F \rangle < 7\times
10^{\hatcurPPfluxavgdim}$\,\ergscmsq{}. The color of the
points encodes metallicity, going
from blue for $\feh{}\leq 0$ to red for $\feh\geq 0.5$ as indicated
in the figure.
\label{fig:MR}}
\end{figure}

We have presented in this work the discovery of \hatcurb{} by the
HATSouth survey, which combines data from three stations around the
globe with close to optimal longitude separations in order to detect
transits. For \hatcurb{} all of the six HATSouth units contributed to
the observations due to it being located in the overlap of two fields
that were monitored by the network, resulting in a substantial 33,212
photometric measurements.

The most remarkable properties of the \hatcur{} system are the high
metallicity of its star and the modest radius of \hatcurb{} given its
mass of $\mpl \approx \hatcurPPmshort$\,\mjup, or equivalently its
high density for its mass.  The metallicity of \hatcur{} makes it more
metal rich than $>$99\% of all planet hosting stars known to date.
We show in the top panel of Figure~\ref{fig:MR} how \hatcurb{}
compares to all other known TEPs on a mass-radius diagram, while the
lower panel shows only planets with irradiation levels $\lesssim$ that
of \hatcurb{}, i.e. $\langle F \rangle < 7\times
10^{\hatcurPPfluxavgdim}$\,\ergscmsq{}. It is immediately clear that
\hatcurb{} is among the most compact known exoplanets in the $\approx$
1--2 Jupiter mass range, even when restricted to systems that have
modest irradiation levels.  A straightforward inference from this fact
is that \hatcurb{} is likely to have a significant amount of heavy
elements. A comparison with the models of
\citet{fortney:2007:core_models} suggests a core mass of $\approx$ 75
\mearth{}.
Its irradiation level of $\langle F \rangle=\hatcurPPfluxavg$
$10^{\hatcurPPfluxavgdim}$\,\ergscmsq{} is near the level below which
planets are found to be not inflated, and thus the uncertainties
introduced by any model assumptions as to the relation between the
incident flux and the mechanism that produces inflated radii are
expected to be minor.
While the uncertainties in the models do not allow us to make firm
conclusions as to the precise value of the core mass, or indeed if
there is a core at all as the heavy elements could be fully mixed in
an envelope, the conclusion that \hatcurb{} is likely to have a
significant content of heavy elements can be obtained just by its
position close to the lower envelope in the mass-radius diagram of the
population of Jupiter mass planets with modest irradiation levels.

The significant heavy element content of \hatcurb{} may be related to
the high metallicity of the host (\feh$=\hatcurSMEzfeh$).
\citet{miller:2011:cores} found that for transiting planets with low
irradiation levels, systems around stars with high stellar metallicity
tend to have a higher mass of heavy elements.
The trend presented in \citet{miller:2011:cores} has a large scatter
though.  In the bottom panel of Figure~\ref{fig:MR} the redder points
represent more metal-rich systems. Although none of the redder points
appear close to the upper envelope of the distribution, the most
compact systems are not exclusively the most metal-rich ones.
The heavy element content of a planet probably depends on many
factors, including where in the protoplanetary disk it formed, when it
formed relative to the disk lifetime, and its detailed migration
history, for example if significant migration happened while the disk
is still present. Such factors could give rise to a large scatter in
any correlation between planet and host star metal fractions.
There also appear to be strong exceptions that call into question the
existence of a very definite correlation between heavy element content
and \feh{}, such as the very compact CoRoT-13b
\citep{cabrera:2010:corot13}, which has a host star with \feh{}
$\approx 0$ and a heavy element content $\approx 200$ \mearth{}. Even
though its irradiation level is higher than those of the sample
considered by \citet{miller:2011:cores}, a higher irradiation should
result, if anything, in an inflated radius. CoRoT-13 does have a large
overabundance of Li (+1.45 dex), so it may be that detailed abundances
of host stars are more relevant than overall metallicity when
considering correlations with the heavy element content of
planets. The tenuousness of the correlation of heavy element content
with \feh{} is illustrated by the recent analysis of
\cite{zhou:2014:hats5} that fitted for linear dependencies of hot
Jupiter radii on both planetary equilibrium temperature and \feh{},
finding that there is no statistically significant correlation with
\feh{} as would be expected if higher \feh{} led tightly to higher
heavy element content.
As transiting surveys such as HATSouth uncover more systems residing
in stars of extreme iron element content such as \hatcurb{} we will be
able to assess with ever more confidence the role of stellar
metallicity in determining the heavy element content of exoplanets.

%


\acknowledgements 

A.J. thanks Didier Queloz for discussions that provided important
insights on the measurement of precision radial velocities with
CORALIE. A.J. also enjoyed many useful discussions on the topic of
radial velocity measurements with James Jenkins.
Development of the HATSouth project was funded by NSF MRI grant
NSF/AST-0723074, operations have been supported by NASA grant NNX09AB29G
and NNX12AH91H, and
follow-up observations receive partial support from grant
NSF/AST-1108686.
A.J. acknowledges support from FONDECYT project 1130857, BASAL CATA
PFB-06, and projects IC120009 ``Millennium Institute of Astrophysics
(MAS)'' and P10-022-F of the Millennium Science Initiative, Chilean
Ministry of Economy.  R.B. and N.E. are supported by
CONICYT-PCHA/Doctorado Nacional. R.B. acknowledges additional support
from Nucleus P10-022-F of the Millennium Science Initiative, Chilean
Ministry of Economy.  V.S.\ acknowledges support form BASAL CATA
PFB-06.  M.R.\ acknowledges support from FONDECYT postdoctoral
fellowship 3120097.
This work is based on observations made with ESO telescopes at the La
Silla Observatory under programme IDs 089.C-0440(A), 090.D-0061(A) and
090.A-9012(A).
Australian access to the Magellan Telescopes was supported through the
National Collaborative Research Infrastructure Strategy of the
Australian Federal Government.  
This paper also uses observations obtained with facilities of the Las
Cumbres Observatory Global Telescope, with the Euler telescope in La
Silla through time awarded by the CNTAC under program CN2012A-62 and
with the Swope telescope in Las Campanas Observatory through time
awarded under program CN2013A-171.
Work at the Australian National University is supported by ARC Laureate
Fellowship Grant FL0992131.
We acknowledge the use of the AAVSO Photometric All-Sky Survey
(APASS), funded by the Robert Martin Ayers Sciences Fund, NASA's
Astrophysics Data System Bibliographic Services, and the SIMBAD
database, operated at CDS, Strasbourg, France.
Operations at the MPG/ESO 2.2\,m Telescope are jointly performed by
the Max Planck Gesellschaft and the European Southern Observatory. We
thank Timo Anguita and R\'egis Lachaume for their technical assistance
during the observations at the MPG/ESO 2.2\,m Telescope.


\appendix
\section{Adopted Data Reduction and Analysis Procedures for CORALIE}

The CORALIE instrument is a fibre-fed echelle spectrograph mounted on
the 1.2m Euler Swiss Telescope at the ESO La Silla Observatory
\citep{queloz:2001:coralie}. The resolving power of CORALIE is now
$R\approx 60000$ after a refurbishment of the instrument performed in
2007. The full width at half-maximum of an unresolved spectral line is
sampled by $\approx$ 3 pixels of the 2k$\times$2k CCD detector.
The spectrograph contains two fibers, an object fiber that during
science observations is illuminated by the target and a comparison
fiber that is simultaneously illuminated either by a wavelength
reference lamp to measure instrumental velocity drifts, or by the sky.
The echelle grating divides the object spectrum in 71 useful orders which are
distributed over the whole CCD, giving spectral coverage of $\approx$
3000 \ang{} in a single exposure. Only the 48 bluest 

The CORALIE reduction procedure adopted by HATSouth observations was
briefly described in \citet{penev:2013:hats1}. Here we describe the
pipeline in more detail. The core of the pipeline was written in
Python, while some computationally intensive tasks were written in C.
The pipeline was developed following a four step structure, which can
be identified as: pre-processing, extraction, wavelength calibration,
and post-processing. In the following paragraphs we describe each step
separately.

\subsection{Pre-Proccessing}

The pipeline starts with the classification of the different image
frames present in a directory according to the FITS header
information. The different frames our CORALIE pipeline uses are bias,
quartz (fiber flats), ThAr frames (object or comparison fiber
illuminated by the wavelength reference lamp) and science frames (in
our case, with the object fiber illuminated by a star and the
comparison fiber by the wavelength reference lamp).  Once the
classification is done, master bias and flat frames are constructed by
median combining biases and fiber flats. The master bias is used to
subtract the bias structure while the master flats are used to trace all
echelle orders for both the object and comparison fibres. The latter
step relies on order finding and tracing algorithms, which we now
describe.

The order finding algorithm starts by median filtering the central
column of the master flat with a 45-pixel window. Then, a robust mean
and standard-deviation of this baseline-subtracted column is obtained
and an order is considered as detected if a given pixel and two
consecutive ones have a signal of more than 5 standard deviations from
the mean. The positions of these consecutive pixels are saved and then
their signals are cross-correlated with a Gaussian in order to find
robust pixel positions for the center of the profile of each order at
the central column.  The process is repeated for adjacent columns, and
the centers for each order are finally fitted with a fourth degree
polynomial to determine the trace for each order.

\subsection{Extraction}

Once orders have been traced using the master flat frames we proceed
to extract the flux of the science and ThAr frames. For the latter we
perform a simple extraction, i.e. just summing the flux in an a region
$\pm 3$ pixels around the trace. The same is true of the comparison
fiber of the science frames which is illuminated by the ThAr lamp. For
the object fiber in the science frames we perform optimal extraction
for distorted spectra as presented in \citet{marsh:1989:optimal}.

\subsection{Wavelength Calibration}

Wavelength calibration is computed based on the emission lines of the
spectrum of a Thorium-Argon (ThAr) lamp. The pipeline only identifies
lines contained in the line list presented in \citet{lovis:2007:thar}.
Reference files are created once for every echelle order. In these
reference files each line is associated with an approximate pixel
position along the trace.  The pipeline then fits Gaussians around
these zones after determining an offset in pixel space between the
positions in the reference file and the lines of a given echelle order
in the frame under processing via cross-correlation.  The Gaussian
means give precise pixel positions of the lines, allowing us to
determine the relation between wavelength, pixel and echelle order for
each emission line. In the case of blended lines, multiple Gaussians
are fitted. An individual wavelength solution is first computed for
each echelle order and ThAr lines that deviate significantly from the
fit are rejected iteratively until the RMS is below 75 \ms. Then a
global wavelength solution in the form of an expansion of the grating
equation \citep[see \S2.6 in][]{baranne:1996:elodie} using Chebyshev
polynomials is fitted and more lines are iteratively rejected until
the RMS of the global solution is below 100 \ms.  Our global
wavelength solution takes the following form

\begin{equation}
\lambda(x,m) = \frac{1}{m}\sum_{i=0}^{3}\sum_{j=0}^{4} a_{ij} c^i(m) c^j(x)
\end{equation}

\noindent where $x$ and $m$ refers to pixel value and echelle order
number respectively, $c^n$ denotes the Chebyshev polynomial of order
$n$, and $a_{ij}$ are the coefficients that are fitted for to obtain the
wavelength solution. The echelle order numbers in CORALIE range from
$m_i=89$ (reddest) to $m_f=159$ (bluest).

Wavelength solutions are first computed for object and comparison
fibers of all ThAr calibration frames and each science frame is
associated to the closest ThAr frame in time. 
Then, the pipeline determines the instrumental velocity shift $\delta
v$ between the science frame and the associated ThAr frame by fitting
a velocity-shifted version of the wavelength solution of the
comparison fiber of the associated ThAr frame to the comparison fiber
of the science frame.
Finally, the object fiber wavelength solution of the associated ThAr
frame is applied to the science object fiber shifted by the
instrumental velocity shift $\delta
v$ computed with the comparison fibers. In general,
the instrumental shifts are smaller than 10 \ms{} on timescales of
$\sim$ 1 hour.

\subsection{Post-Processing}

After obtaining the wavelength calibrated spectrum of a star, the
pipeline determines the barycentric correction and applies it to the
wavelength solution using routines from the Jet Propulsion
Laboratory ephemerides (JPLephem) package.
We then apply a fast and robust stellar parameter determination
to each stellar spectrum.
In order to estimate the stellar parameters (\teffstar, \logg, \feh,
\vsini) we cross-correlate the observed spectrum against a grid of
synthetic spectra of late type stars from \citet{coelho:2005}
convolved to the resolution of CORALIE and a set of \vsini{} values
given by $\{0,2.5, 5, 7.5, 10, 15,\ldots,45,50\}$ \kms.  First a set
of atmospheric parameters is determined as a starting point by
ignoring any rotation and searching for the model that produces the
highest cross-correlation using only the Mg triplet region. We use a
coarse grid of models for this initial step ($\Delta \teffstar = 1000$
K, $\Delta \logg = 1.0$, $\Delta \feh = 1.0$). A new cross-correlation
function (CCF) between the observed spectrum and the model with the
starting point parameters is then computed using a wider spectral
region (4800 \ang -- 6200 \ang). From this CCF we estimate RV and
\vsini{} values using the peak position and width of the CCF,
respectively. New \teffstar, \logg{} and \feh{} values are then
estimated by fixing \vsini{} to the closest value present in our grid
and searching for the set of parameters that gives the highest
cross-correlation, and the new CCF is used to estimate new values of
RV and \vsini.  This procedure is continued until convergence on
stellar parameters is reached.
For high S/N spectra, uncertainties of this procedure are typically
200 K in \teffstar, 0.3 dex in \logg, 0.2 dex in \feh{} and 2 \kms{}
in \vsini. These uncertainties were estimated from observations of
stars with known stellar parameters from the literature.  Results of
our stellar parameter estimation procedure are obtained in about 2
minutes using a standard laptop
Fast and robust atmospheric parameters are estimated in order to make
efficient use of the available observing time because in this way some
false positives or troublesome systems, such as fast rotators and
giants, can be identified on the fly during the follow-up process.

\begin{figure}[!ht]
\plotone{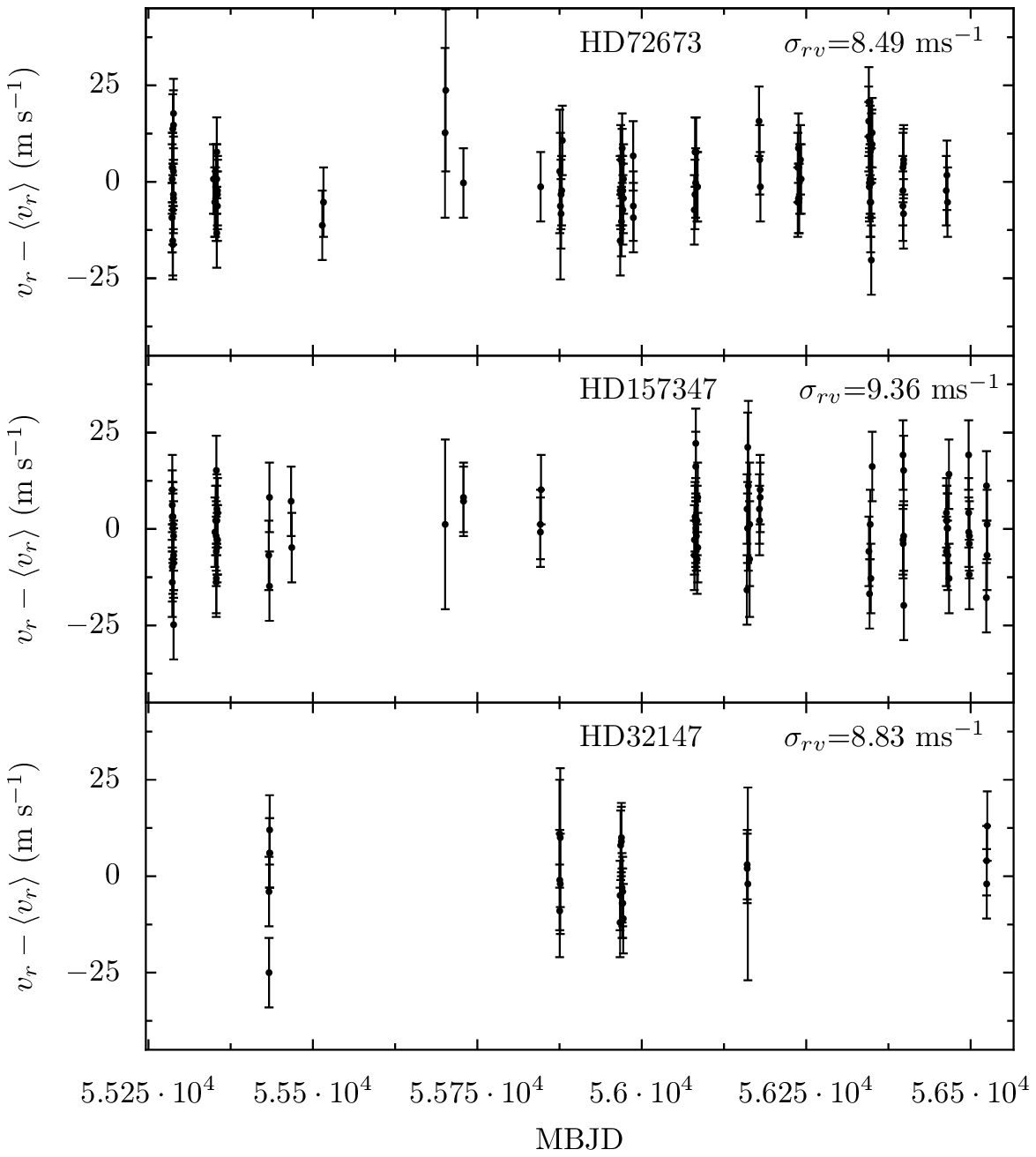}
\caption{
       Radial velocity $v_r$ time series for the 3 RV standard stars
       we have monitored with CORALIE. In each panel we show the name of the
       standard and the root mean square deviation around the mean RV
       ($\langle v_r \rangle$). We conclude from these data that our
       long-term stability is $\approx$ 9 \ms.
\label{fig:RVstandards}}
\end{figure}

The radial velocity of the spectrum is computed using the
cross-correlation technique against a binary mask
\citep{baranne:1996:elodie} which depends on the spectral type. The
available masks are for G2, K5 and M5, and are the same ones used by
the HARPS \citep{mayor:2003:harps} data reduction system but with the
transparent regions made wider given that the resolution of CORALIE is
$\approx$ 0.5 times that of HARPS.
If a target is being observed for the first time, first a query to the
SIMBAD database is made to see if the spectral type is known and
otherwise our estimate of \teffstar{} is used to assign the mask to
calculate the radial velocity.
Future observations of the same target will use always this same mask
to compute the radial velocity.
The CCF is computed for each echelle order and then a weighted sum is
applied to get the final CCF.  Weights are given by the mean
S/N on the continuum for each echelle order.
A simple Gaussian is fitted to the CCF, and the mean is taken as the
RV of the spectrum.  The pipeline also infers the presence of
moonlight contamination in the CCF.  If there is a secondary peak on
the CCF due to the moon, the pipeline computes the position and width
of this peak and fits for the intensity to correct the RV
determination.  Bisector spans are also measured using the CCF peak as
described in \citet{queloz:2001:hd166}.  Uncertainties on RV and
bisector span are determined from the width of the CCF and the mean
S/N close to the Mg triplet zone using empirical scaling relations
\citep[as in, e.g.,][]{queloz:1995:ccf} whose parameters are
determined using Monte Carlo simulations where Gaussian noise is
artificially added to high S/N spectra. The exact equations used to
estimate the errors are

\begin{equation}
\sigma_{\rm RV} = b + \frac{a( 1.6 + 0.2  \sigma_{\rm ccf} ) }{ {\rm SN}_{5130}}
\end{equation}
\begin{equation}
\sigma_{\rm BS} = \frac{d}{{\rm SN}_{5130}} + c
\end{equation}

\noindent where $a,b,c,d$ are the coefficients obtained via the Monte
Carlo simulations and depend on the applied mask, ${\rm SN}_{5130}$
is the continuum S/N at 5130 \ang{} and $\sigma_{\rm ccf}$ is the
dispersion of the Gaussian fit to the CCF. For illustration, the
values of the coefficients for the G2 mask are $a=0.06544, b=0.00146,
c=0.00181$ and $d=0.24416$. If the estimated uncertainty $\sigma_{\rm
  RV} < 9\, \ms$ we assign it a value of 9 \ms{} given that the
long-term radial velocity precision we achieve is $\approx$ 9 \ms{}
over a timescale of several years. We determined this value by
monitoring 3 radial velocity standard stars (HD72673, HD157347,
HD32147) starting around 2010. Figure~\ref{fig:RVstandards} shows the
radial velocity time series for all the CORALIE measurements we have
taken of the standards, whose spectra have ${\rm SN}_{5130}$ in the
range $\sim$ 50--150.

\bibliographystyle{apj}
\bibliography{refs}

\end{document}